\documentclass[11pt]{article}
\pdfoutput = 1
\usepackage[utf8]{inputenc}

\usepackage[top=1in,bottom=1in,left=1in,right=1in]{geometry}
\makeatletter \g@addto@macro\@floatboxreset\centering \makeatother
\usepackage{color}
\definecolor{darkgreen}{rgb}{0,0.5,0}
\definecolor{darkblue}{rgb}{0,0,0.6}
\definecolor{purple}{rgb}{0.4,.2,0.7}
\usepackage[noEucal]{main}
\usepackage{aas_macros,empheq,fancybox,graphicx,multirow}
\usepackage{aas_macros}
\usepackage{ulem}
\usepackage{dsfont}
\usepackage{esint}
\usepackage{float}
\usepackage{appendix}
\usepackage{mathabx}
\usepackage{hhline}
\usepackage{tensor} 
\usepackage{footmisc}
\usepackage{float}
\usepackage{caption}
\begin{document}

\thispagestyle{empty}

\begin{center}
    ~
    \vskip10mm

     {\LARGE  {\textsc{Integrable Field Theories and Their CCFT Duals}}}
    \vskip10mm
    
Daniel Kapec$^{a,b}$ and  Adam Tropper$^{b}$ \\
    \vskip1em
    {\it
        $^a$ Center of Mathematical Sciences and Applications, Harvard University, Cambridge, Massachusetts 02138, USA\\ \vskip1mm
        $^b$ Center for the Fundamental Laws of Nature,
Harvard University, Cambridge, Massachusetts 02138, USA\\ \vskip1mm
         \vskip1mm
    }
    \vskip5mm
    \tt{danielkapec@fas.harvard.edu, adam$\_$tropper@g.harvard.edu}
\end{center}
\vspace{10mm}

\begin{abstract}
\noindent

We compute the Mellin transforms of various two-dimensional integrable $S$-matrices, providing the first explicit, non-perturbative realizations of  celestial CFT. In two dimensions, the Mellin transform is simply the Fourier transform in rapidity space, and the ``celestial correlator'' has no position dependence. The simplified setting allows us to study the analytic properties of CCFT correlators exactly as a function of the conformal dimensions. We find that the correlators exist as real distributions of the conformal weights, with asymptotics controlled by the mass spectrum and three-point couplings of the model. Coupling these models to a flat space limit of JT gravity preserves integrability and dresses the amplitudes by a rapidly varying gravitational phase. We find that the coupling to gravity smooths out certain singular aspects of the Mellin-transformed correlators. 

\end{abstract}
\pagebreak

\setcounter{tocdepth}{2}
{\hypersetup{linkcolor=black}
\small
\tableofcontents
}

\section{Introduction}
Despite our detailed understanding of the holographic principle in asymptotically AdS spacetimes, flat space holography remains an underdeveloped subject. In asymptotically flat spacetimes, the black hole entropy still scales like the area and the gravitational Hamiltonian is still a boundary term. It is natural to expect that some form of holography applies to quantum gravity with flat asymptotics, but the precise statement of the correspondence has so far remained elusive. The sheer number of open questions makes the subject ripe for exploration, but so far progress has been slow. 

Most attempts at flat holography to date seek to use null infinity  $(\mathcal{I}^{\pm})$ as the gravitational hologram (see \cite{Susskind:1998vk,deBoer:2003vf,Mann:2005yr,Dappiaggi:2005ci,Barnich:2010eb,Bagchi:2010zz,Banks:2014iha,Pasterski:2017kqt} for some early attempts and representative perspectives). The approaches with the most concrete results take an explicitly bottom-up approach, translating known bulk results into a seemingly holographic formalism. In this sense these approaches cannot be wrong. The question is simply whether or not they are useful.

Celestial CFT (CCFT) is one such bottom-up approach to quantum gravity in flat space. In short, it seeks to construct an independent definition of the $S$-matrix intrinsic to $\mathcal{I}^\pm$, with no reference to bulk propagation. It is close in spirit to the AdS/CFT correspondence, whose conformal correlators provide an independent and nonperturbative definition of boundary observables in AdS quantum gravity. The formalism relies on the fact that the $d+2$ dimensional Lorentz group ${SO(d+1,1)}$ is isomorphic to the $d$-dimensional Euclidean conformal group, which in turn guarantees that the $S$-matrix shares some, but certainly not all, of the properties of a correlation function in CFT$_d$. Since the geometric locus $\mathcal{I}$ is also the momentum space locus of unbound null geodesics, the celestial correlators have operator insertions where the particles enter or exit the spacetime 
\begin{equation*}
 \langle p_1 , \cdots , p_m | p_{m+1} , \cdots , p_n \rangle \quad \sim \quad \langle \mathcal{O}_1(\omega_1,x_1) \cdots \mathcal{O}_n(\omega_n,x_n) \rangle \; .
\end{equation*}
In a basis of boost eigenstates these matrix elements transform like conformal correlators in CFT$_d$, and ultimately one would like an intrinsically $d$-dimensional method for computing them. 

Certain structural results shared by all CCFT's are available \cite{Kapec:2017gsg,Kapec:2022hih}, but they are limited in scope and primarily controlled by symmetries (although see \cite{Kapec:2022axw}). Beyond these universal results, most work on the subject takes as input perturbative scattering amplitudes and returns Mellin transformed correlators and OPE coefficients (see for example \cite{Pasterski:2017ylz,Schreiber:2017jsr,Stieberger:2018edy,Nandan:2019jas,Pate:2019lpp,Gonzalez:2020tpi,Banerjee:2020vnt,Arkani-Hamed:2020gyp,Guevara:2021abz,Fan:2021isc,Atanasov:2021cje,Himwich:2021dau,Fan:2021pbp,Adamo:2021zpw,Hu:2022syq,Garcia-Sepulveda:2022lga,Ren:2022sws,Bhardwaj:2022anh,Stieberger:2022zyk} and references therein). These results are obtained order by order in perturbation theory, and as a result, the non-perturbative defining properties of CCFT are not known. We lack an axiomatic approach or intrinsic definition.

This gap in the foundations of CCFT stems in part from the lack of a complete set of defining relations for the non-perturbative $S$-matrix in asymptotically flat space. Beyond the standard inputs of locality and unitarity and the assumption of analyticity, most of the known consistency conditions are derived from perturbation theory. The complete set is simply unknown, and what to expect when gravity and black holes are included is even less clear.

Given the extent of these uncertainties, a top-down construction with no question marks is clearly desirable. Well-defined CCFTs are believed to arise from strings propagating in asymptotically flat space, but the non-perturbative properties of those models are still not understood and they are certainly not the simplest place to start. There has been interesting recent work \cite{Costello:2022wso,Costello:2022jpg} on somewhat exotic 4d models with explicit CCFT duals, along with some work on self-dual gravity \cite{Ball:2021tmb}, which may prove to be the simplest tractable model of celestial CFT. In this paper, we consider another class of $S$-matrices which are known exactly and whose Mellin transforms we can compute explicitly: the integrable $S$-matrices in two dimensions. 

Integrable quantum field theories are completely non-generic, and it is reasonable to question how much one might learn about CCFT by studying them. The existence of infinitely many conserved quantities puts severe constraints on the form of the scattering amplitudes in these theories, but many of the peculiar properties that distinguish CCFT$_d$ from garden-variety Euclidean CFT$_d$ are still present in these dramatically simplified models. As we will see, the conformal correlators still exhibit complicated analytic structure as a function of the operator dimensions, and the theories can be coupled to gravity in a simple but illuminating way. The calculations in this paper provide a rigorous glimpse into the possible behaviors of CCFT correlation functions, but we make no claims regarding genericity. It would certainly be worthwhile to extend the analysis to non-integrable two-dimensional models in order to understand which conclusions hold more broadly, and we hope to return to this question in the future.

Since the bulk $S$-matrix is two dimensional in our examples, the CCFT$_0$ correlators have no position dependence and the Mellin transform takes the same form for massive and massless particles. In fact, the Mellin transform with conformal weights on the continuous series simply reduces to a Fourier transform in rapidity space which can be evaluated explicitly using standard techniques in complex analysis. This simplified setting allows us to make very general statements regarding the analytic structure of the conformal correlators as functions of the operator dimensions. In particular, we find that the correlators are generally distributional, with asymptotics controlled by the mass spectrum and fusion coefficients of the quantum field theory. Although gravity does not have propagating degrees of freedom in two dimensions, we find that coupling these $S$-matrices to flat space JT gravity alters the celestial correlators in interesting ways. For instance, the coupling to gravity smooths out certain singular aspects of the non-gravitational correlators. 

Although we obtain exact expressions for the CCFT$_0$ ``correlation numbers,'' we stop short of constructing a celestial dual in the sense that we lack a (zero-dimensional) path integral from which the Mellin-space correlators can be derived. We hope to revisit this question, as well as the decomposition of the correlators under operator exchange, in later work. Since there is no guarantee that celestial correlators arise from a Lagrangian, the complete list of correlators presented here is equivalent to a definition of the model. 

While this paper was in preparation, a related work \cite{Duary:2022onm} appeared which also considers the Mellin transform of a particular 2d $S$-matrix. This paper works perturbatively and has minimal overlap with our results, but the model considered (Sinh-Gordon) is also integrable and could be treated non-perturbatively within our more general framework. 

The outline of this paper is as follows. In section \ref{sec:Kinematics} we review the celestial CFT$_d$ formalism in the special case $d=0$. Section \ref{sec:IntSmat} reviews relevant properties of integrable theories and translates them into properties of the celestial dual. 
In section \ref{sec:Diagonal}, we explicitly compute the celestial amplitudes for a particular class of integrable theories solved via the S-matrix bootstrap and discuss their general features. Section \ref{sec:grav} considers the effect of coupling these integrable field theories to a simple model of 2d gravity. We conclude with a discussion of open questions and future work in section \ref{sec:Discussion}.

\section{Two-dimensional Kinematics and CCFT$_0$}\label{sec:Kinematics}

We will consider quantum field theories defined on $\mathbb{R}^{1,1}$. In two dimensions, the Lorentz group $SO(1,1)$ consists only of boosts, and 
the only ``celestial conformal generator'' is the dilation operator $D=M_{10}$.
A generic massive momentum can be written\footnote{Since the null cone in two dimensions is disconnected, the parameterization for massless particles takes a slightly different form. One would use $p=\omega_R\hat{q}$ and $p=\omega_L n$ for left and right moving momenta and Mellin-transform with respect to those variables. }
\begin{equation}
\begin{split}
p^\mu(\omega) = \omega {\hat p}^\mu ( \omega  ) \; , \qquad {\hat p}^\mu(\omega) = {\hat q}^\mu + ( m^2 / \omega^2 )  n^\mu \; , 
\end{split}
\end{equation}
where ${\hat q}$ and $n$ correspond to left and right moving branches of the null cone
\begin{equation}
\begin{split}
{\hat q}^\mu(x) = \frac{1}{2} ( 1 ,  1) \; , \qquad n^\mu = \frac{1}{2} ( 1  , - 1 )  \;  , \qquad \hat{q}\cdot n=-\frac12 \; . 
\end{split}
\end{equation}
Lorentz invariance fixes the transformation properties for both massless and massive particles
\begin{equation}
\label{eq:trans}
[ \mathcal{O}(\omega) , D ] = -i  \omega \partial_\omega  \mathcal{O}(\omega) \;  .
\end{equation}
The Mellin transform is therefore the same for both gapped and gapless states (in contrast to the higher dimensional case, where the Mellin transform for massive particles involves an extra integral over a bulk-to-boundary propagator):
\begin{equation}\label{eq:Mellin}
\widehat{\mathcal{O}}(\hat{\Delta}) = \int_{\mathcal{C}} d\o \o^{\hat{\Delta}-1} \mathcal{O}(\omega) \; , \qquad [\widehat{\mathcal{O}}(\hat{\Delta}), D]=i\hat{\Delta}\widehat{\mathcal{O}}(\hat{\Delta}) \; .
\end{equation}
There are different possible choices for the contour of integration and admissible values of $\hat{\Delta}$ in this formula. The most common choice is a non-compact contour with $\hat{\Delta}$'s on the continuous series
\begin{equation}\label{eq:continuous}
\widehat{\mathcal{O}}(\hat{\Delta}) = \int_{0}^\infty d\o \o^{\hat{\Delta}-1} \mathcal{O}(\omega) \; , \qquad    \hat{\Delta}=\frac{d}{2}+i\Delta \; ,
\end{equation}
which in the case at hand is simply  
\begin{equation}
    \hat{\Delta}=i\Delta \; .
\end{equation} 
This choice notably excludes the integer scaling dimensions of crucial operators like the stress tensor and conserved currents, but these operators do not appear in CCFT$_0$ (there are no gravitons or gluons in two dimensions) so the distinction will not be important. More importantly, the formula \eqref{eq:continuous} depends on the behavior of scattering amplitudes at arbitrarily large energies and its meaning is murky within the standard effective field theory framework. In this paper we consider UV complete exact $S$-matrices, so this is not a concern and we adopt the definition
\begin{equation}\label{eq:MellinContinuous}
\widehat{\mathcal{O}}(\Delta) = \int_{0}^{\infty} \frac{d\omega}{\omega} \omega^{i\Delta} \mathcal{O}(\omega) \;. 
\end{equation}
In two dimensions it is useful to change coordinates to rapidity space $m/\omega=e^{-\theta}$, in which case
\begin{equation}
    p=m(\cosh \theta, \sinh \theta)=m(e^\theta \hat{q} + e^{-\theta} n) \; . 
\end{equation}
The variable $\theta$ is known as the rapidity, and it is the variable conjugate to dilations (boosts)
\begin{equation}
    D=-i\partial_\theta \; .
\end{equation}
The boost eigenstates are then simply Fourier transforms with respect to rapidity\footnote{We normalize our operators without the conventional factor of $m^{i\Delta}$.} 
\begin{equation}
 \widehat{\mathcal{O}}(\Delta)=   \int_{-\infty}^{\infty} d\theta e^{i\Delta \theta} \mathcal{O}(\theta) \; .
\end{equation}
Due to the low dimensionality, the ``celestial correlation functions'' are just functions of $\Delta$
\begin{equation}\label{eq:corrftns}
    \widetilde{S}(\Delta_1,\cdots \Delta_n)= \langle \widehat{\mathcal{O}}(\Delta_1) \cdots \widehat{\mathcal{O}}(\Delta_n)  \rangle  \; .
\end{equation}
When $d>0$, celestial CFT$_d$ correlators have complicated dependence both on $\Delta$ and on the transverse coordinates $x^a$. The quantities \eqref{eq:corrftns} are of interest precisely because they can be used to study the $\Delta$ dependence of the celestial  amplitudes while avoiding the complicated analysis associated to position dependence. The simplification \eqref{eq:corrftns} is similar in spirit to studying matrix integrals as toy models of gauge theory.

\section{CCFT Dual to Integrable Theories I: The General Case}
\label{sec:IntSmat}
Integrable $S$-matrices enjoy a number of remarkable properties not shared by generic quantum field theories. Here we review these properties and recast them as constraints on the CCFT dual.

\subsection{Elastic Scattering}

The existence of infinitely many conserved quantities in integrable theories places severe constraints on the $S$-matrix. In more than two dimensions, higher-spin conserved charges lead to a trivial $S$-matrix \cite{Coleman:1967ad}, but in two dimensions it is still possible to have nontrivial scattering processes because of the restricted kinematics. In this case, integrability guarantees that all scattering is elastic: there is no particle production and the sets of initial and final momenta are the same. The number of particles with fixed mass also cannot change during the scattering process, so the rapidities agree
\begin{equation}\label{eq:RapidCons}
   \{ \theta_i \}_{\text{in}} = \{\theta_i'\}_{\text{out}} \qquad \qquad  i = 1,\dots,n \; .
\end{equation}
In what follows primes denote out-state quantities and we suppress the \textit{in} and \textit{out} labels. Lorentz (boost) invariance implies that the $S$-matrix only depends on differences of rapidities $\theta_{ij} = \theta_i - \theta_j$, and we will choose $\theta_{12},\theta_{13},...,\theta_{1n}$ as a basis for the kinematics. Scattering amplitudes in these models can only be nonzero on the support of the constraint \eqref{eq:RapidCons}. They come with a product of delta functions that enforce the conservation of individual rapidities (and masses)
\begin{equation}\label{eq:elastic}
    S_{a_1 \cdots a_n}^{b_1 \cdots b_n}(\theta_1,...,\theta_n;\theta'_1,...,\theta'_{n}) =  S_{a_1 \cdots a_n}^{b_1 \cdots b_n}(\theta_{12},...,\theta_{1n}) \prod_{i=1}^n \delta(\theta_i - \theta'_i) ~ + ~ \text{permutations}\; .
\end{equation}
The additional permutations are related to non-diagonal scattering and are only relevant when there are degeneracies in the mass spectrum. 
To obtain the celestial correlator, we take the Fourier transform of both sides of \eqref{eq:elastic}
\begin{equation}
    \begin{split}
        \widetilde{S}^{b_1 \cdots b_n}_{a_1 \cdots a_n}(\Delta_i;\Delta_i') &= \delta\big(\Delta_1 + \Delta_1' + \cdots + \Delta_n + \Delta_n'\big) \bigg[\widetilde{M}_{a_1 \cdots a_n}^{b_1 \cdots b_n}(\Delta_i;\Delta_i') + \text{permutations}\bigg] \; .\\
    \end{split}
    \label{eqn:S-matrixdeltaconstraint}
\end{equation}
The celestial $S$-matrix is only supported on a delta function, $\delta\big(\sum_{i=1}^n (\Delta_i + \Delta_i')\big)$, which arises from the integral over the collective mode $\sum_i \theta_i$. The presence of this delta function is simply the Ward identity associated to boost symmetry in the bulk (dilation symmetry in the boundary).\footnote{In higher dimensions, $[O(\omega,x),D]=-i(\omega \partial_\omega -x^a\partial_a)$, so the Ward identity for insertions of $D$ involves extra derivatives. In two dimensions there is no position derivative, so switching to the Mellin basis completely diagonalizes the operator $D$ and boost weight is automatically conserved.}
We will refer to $\widetilde{M}^{b_1 \cdots b_n}_{a_1 \cdots a_n}(\Delta_i;\Delta_i')$ as the $n \to n$ celestial amplitude, keeping in mind that it is always accompanied by an overall boost-conserving delta function.

Note that for the celestial correlators obtained from integrable QFTs, it is crucial to distinguish between conformal primaries associated to the \textit{in} states versus those arising from the \textit{out} states. Because of the extra symmetries in these models, the $S$-matrix is not analytic on the support of the overall momentum-conserving delta function. This is apparent in \eqref{eq:elastic}, which has many more delta functions than required by momentum conservation. This singular behavior means that naive applications of crossing symmetry fail. For instance, the existence of a $3\to 3$ amplitude does not imply the existence of a $2\to 4$ amplitude. Any non-vanishing celestial correlator with $2n$ operator insertions involves $n$ \textit{in}-operators and $n$ \textit{out}-operators.

To simplify notation, it is convenient to define $\widehat{\Delta}_i$ as the sum of the conformal weights for ingoing and outgoing particles of the same rapidity
\begin{equation}
    \widehat{\Delta}_i = \Delta_i + \Delta_i' \; .
\end{equation}
In this notation, the celestial amplitude is simply an $(n-1)$-dimensional Fourier transform of the rapidity space amplitude\footnote{In non-integrable models there are additional branch-cuts in the rapidity plane associated to particle production, and the definition of \eqref{eqn:NtoNCelestial} may require the choice of a contour around the cuts. }
\begin{equation}\label{eqn:NtoNCelestial}
    \widetilde{M}^{b_1 \cdots b_n}_{a_1 \cdots a_n}(\widehat{\Delta}_2,...,\widehat{\Delta}_n) = \int_{-\infty}^\infty d\theta_{12} \cdots d \theta_{1n} ~ e^{-i(\widehat{\Delta}_2 \theta_{12} + \cdots + \widehat{\Delta}_n \theta_{1n})} S_{a_1 \cdots a_n}^{b_1 \cdots b_n}(\theta_{12},...,\theta_{1n}) \; .
\end{equation}

\subsection{Factorizability and the Yang-Baxter Equation}

Integrability imposes a second strong constraint on the $S$-matrix that allows one to construct higher-point scattering amplitudes from lower-point matrix elements. In general, $n \to n$ scattering processes can be factorized into a sequence of $n(n-1)/2$ successive $2 \to 2$ scattering processes, so knowledge of the $2\to 2$ amplitudes in the model amounts to a solution of the theory. 

There are many equivalent ways to factorize the higher-point amplitude on lower-point processes, and their equality places strong constraints on the $S$-matrix when there is non-diagonal scattering. This constraint takes the simplest form for the $3\to 3$ amplitude, where it is known as the Yang-Baxter equation. In terms of two-particle $S$-matrices it reads
\begin{equation}
    S_{a_1 a_2}^{b_1 b_2}(\theta_{12})S_{b_1a_3}^{c_1 b_3}(\theta_{13})S_{b_2 b_3}^{c_2 c_3}(\theta_{13} - \theta_{12}) = S^{c_1 c_2 c_3}_{a_1 a_2 a_3}(\theta_{12},\theta_{13}) = S^{b_2 b_3}_{a_2 a_3}(\theta_{13} - \theta_{12}) S_{a_1 b_3}^{b_1 c_3}(\theta_{13})S_{b_1 b_2}^{c_1 c_2}(\theta_{12}) \;.
\end{equation}
The Yang-Baxter equation can be thought of as an associativity condition that ensures that the result for a higher-point amplitude doesn't depend on the choice of factorization.

Since multiplication in rapidity space is convolution in Fourier space, the Yang-Baxter equation imposes the following constraint on integrable celestial amplitudes
\begin{equation}
    \begin{split}
        \widetilde{M}_{a_1 a_2 a_3}^{c_1 c_2 c_3}(\widehat{\Delta}_2,\widehat{\Delta}_3) &= \int_{-\infty}^\infty d\Delta ~  \widetilde{M}_{a_1 a_2}^{b_1 b_2}(\widehat{\Delta}_2 + \Delta) \widetilde{M}_{b_2 b_3}^{c_2 c_3}(\Delta) \widetilde{M}_{b_1 a_3}^{c_1 b_3}(\widehat{\Delta}_3 -\Delta) \\
        &= \int_{-\infty}^\infty d\Delta' ~  \widetilde{M}_{b_1 b_2}^{c_1 c_2}(\widehat{\Delta}_2 + \Delta') \widetilde{M}_{a_2 a_3}^{b_2 b_3}(\Delta') \widetilde{M}_{a_1 b_3}^{b_1 c_3}(\widehat{\Delta}_3 - \Delta') \; ,
    \end{split}
\end{equation}
where the top and bottom lines are the Fourier transform of the left hand side and right hand side of the Yang-Baxter equation, respectively.

\subsection{Real-Analyticity, Unitarity, and Crossing Symmetry}

Factorization implies that the $2 \to 2$ $S$-matrix elements are the fundamental building blocks of rapidity space amplitudes, which means that they are also the fundamental building blocks of higher-point celestial correlators. In the context of integrable CCFT, solving the model therefore amounts to calculating the celestial four-point function. In this section we discuss its basic properties. 

It is typically assumed that the $2$-particle $S$-matrix in an integrable theory $S_{a_1 a_2}^{b_1 b_2}(\theta_{12})$ is a meromorphic function on the complex $\theta_{12}$ plane obeying the axioms of real-analyticity, unitarity, and crossing symmetry \cite{Dorey:1996gd,Paulos:2016but}.
In rapidity space, real analyticity requires that ${S_{a_1 a_2}^{b_1 b_2}(\theta^*_{12}) = S_{a_1 a_2}^{b_1 b_2}(-\theta_{12})^*}$. It follows that the celestial four-point function has no imaginary part. This in turn implies that the $n \to n$ celestial correlator for integrable theories is real
\begin{equation}
    \text{Im} ~  \widetilde{M}_{a_1 \cdots a_n}^{b_1 \cdots b_n}(\widehat{\Delta}_i) = 0 \; .
\end{equation}
Unitarity requires that for real rapidities
\begin{equation}\label{eq:constraint}
    {S_{a_1 a_2}^{b_1 b_2}(\theta_{12})S_{b_1 b_2}^{c_1 c_2}(-\theta_{12}) = \delta^{c_1}_{a_1} \delta^{c_2}_{a_2}} \; .
\end{equation}
Fourier transforming both sides of \eqref{eq:constraint} imposes the following constraint on celestial amplitudes
\begin{equation}
    \frac{1}{(2\pi)^2}\int_{-\infty}^\infty d \Delta ~ \widetilde{M}_{a_1 a_2}^{b_1 b_2}(\Delta) \widetilde{M}_{b_1 b_2}^{c_1 c_2}(\Delta -  \widehat{\Delta}_2) = \delta_{a_1}^{c_1}\delta_{a_2}^{c_2}\hspace{2pt} \delta(\widehat{\Delta}_2) \; .
\end{equation}
The third defining property of $2 \to 2$ rapidity space amplitudes is crossing symmetry which states $S_{a_1 a_2}^{b_1 b_2}(\theta_{12}) = S_{a_1 \mathcal{C}(b_2)}^{b_1 \mathcal{C}(a_2)}(i \pi - \theta_{12})$ where $\mathcal{C}$ is the charge conjugation operator which replaces an external particle with its antiparticle. One may take the Fourier transform of both sides of this expression. However, the result depends non-trivially on the particles appearing in the theory as bound states. A more thorough discussion is provided in appendix \ref{appendix:crossing}.

\subsection{Existence of the $n$-particle Celestial $S$-Matrix}

\label{sec: existence}

As remarked in section \ref{sec:Kinematics}, the Mellin transform as defined by \eqref{eq:MellinContinuous} involves an integration over all energy scales. Since the contour of integration is non-compact, the integral is not guaranteed to converge and the resulting correlation functions might be singular or simply not exist.

For the models considered in this paper, it turns out that the celestial correlator for $n \rightarrow n$ scattering always exists as a tempered distribution. To see this, we first note that an individual $S$-matrix element satisfies $|S_{a_1 a_2}^{b_1 b_2}(\theta_{12})| \leq 1$ by unitarity. Using factorizability of the $n$-particle $S$-matrix into $2$-particle $S$-matrices and unitarity of the constituent $2$-particle $S$-matrices, we deduce that the $n$-particle $S$-matrix satisfies $|S_{a_1...a_n}^{b_1 \cdots b_n}(\theta_{12},...,\theta_{1n})| \leq 1$ as well. Because its modulus is bounded, $S_{a_1...a_n}^{b_1 \cdots b_n}(\theta_{12},...,\theta_{1n})$ is a tempered distribution, so it has a unique Fourier transform which is also a tempered distribution. According to equation \eqref{eqn:NtoNCelestial}, we identify this Fourier transform as the $n \to n$ celestial amplitude $\widetilde{M}_{a_1...a_n}^{b_1 \cdots b_n}(\widehat{\Delta}_i)$.

In general, it is not possible to say more about the celestial correlator. We will see in explicit examples that $\widetilde{M}^{b_1b_2 }_{a_1 a_2}(\widehat{\Delta}_2)$ is generally not bounded and is distributional in nature. Interestingly, coupling to gravity seems to improve this behavior, as we discuss in section \ref{sec:grav}.

\section{CCFT Dual to Integrable Theories II: Diagonal Scattering}\label{sec:Diagonal}

Integrability obviously dramatically simplifies the form of the $S$-matrix, but there is still a high level of complexity among this set of tightly constrained models. An extra simplifying assumption often employed in the context of the $S$-matrix bootstrap is that of \textit{diagonal scattering}.

Elastic scattering guarantees that the masses and rapidities of ingoing particles match those of the outgoing particles, but the $S$-matrix restricted to the $n\to n$ sector can still have non-diagonal components if the model exhibits flavor symmetries and mass degeneracies. For example, particles $a_1$ and $a_2$ will generically scatter into particles $b_1$ and $b_2$ with $a_i \neq b_i$ provided that the mass of $a_i$ matches that of $b_i$. When the $S$-matrix is diagonal, particles $a_1$ and $a_2$ will only scatter into particles $a_1$ and $a_2$ and we can label the rapidity-space $S$-matrices as $S_{ij}(\theta_{12}) = S_{ij}(\theta)$, where $i,j$ now run over the particle species. Although this extra condition is restrictive, it is satisfied by many important examples, including the affine Toda theories \cite{Braden:1989bu}, perturbed coset models \cite{Fateev:1990hy}, and certain limits of Ising field theory \cite{Zamolodchikov:1989fp, Zamolodchikov:2013ama,Hollowood:1989cg,Gabai:2019ryw}. Indeed, the $S$-matrix bootstrap is often carried out with the assumption of diagonal scattering \cite{Dorey:1996gd}. The assumption does exclude important models with global symmetries, such as the $O(N)$ models and Gross-Neveu models \cite{Zamolodchikov:1978xm}. 

As we will see, the celestial dual to the diagonal theories can be worked out in complete detail. The results exhibit many qualitative features that are likely to be relevant in more general settings.

\subsection{The $2 \to 2$ Celestial Amplitude}
\label{sec:2to2celestialAmp}

The most general diagonal $2$-particle $S$-matrix satisfying real-analyticity, unitarity and crossing symmetry obeys the periodicity condition $S_{ij}(\theta) = S_{ij}(\theta + 2\pi i)$. These $S$-matrices have been shown to take the form \cite{Klassen:1989ui}
\begin{equation}\label{eqn: S-matrix expression}
    S_{ij}(\theta) = \prod_{\alpha \; \in \; \Omega_{ij}} f_{\alpha}(\theta) \qquad \qquad f_\alpha(\theta) = \frac{\sinh(\frac{1}{2}(\theta + i \pi \alpha))}{\sinh(\frac{1}{2}(\theta - i \pi \alpha))} \; ,
\end{equation}
where $\Omega_{ij} \subset \mathbb{C}$ is invariant under complex conjugation. We will generally take $\Omega_{ij} \subset \mathbb{R}$ to avoid discussing unstable particles. This $S$-matrix has poles at $\theta = i \pi \alpha$ and one may choose $\alpha \in (0,2)$ by periodicity. For $\alpha \in (0,1)$, the poles indicate the presence of a bound state $B$ of particles $i$ and $j$. If $i \hspace{2pt} \text{Res}\big[S_{ij}(\theta); \theta = i \pi \alpha\big] > 0$, then this bound state occurs for forward channel scattering (rather than a crossed channel) and the particle has mass
\begin{equation}
    m^2_{B} = m_{i}^2 + m_{j}^2 + 2 m_{i} m_{j} \cos(\pi \alpha) \; .
    \label{eqn: bound state mass}
\end{equation}
Moreover, if the pole is simple, then the residue is related to the on-shell three-point coupling constants $f_{Bij}$ of the underlying QFT
\begin{equation}    \label{eqn: bound state coupling}
    (f_{B i j})^2 = i \; \text{Res}\big[S(\theta); \theta = i \pi \alpha\big] \; .
\end{equation}
For $\alpha \in (1,2)$, the poles indicate the presence of an unstable resonance \cite{Dorey:1996gd}.
\begin{figure}
    \centering
    \includegraphics[width = .8 \textwidth]{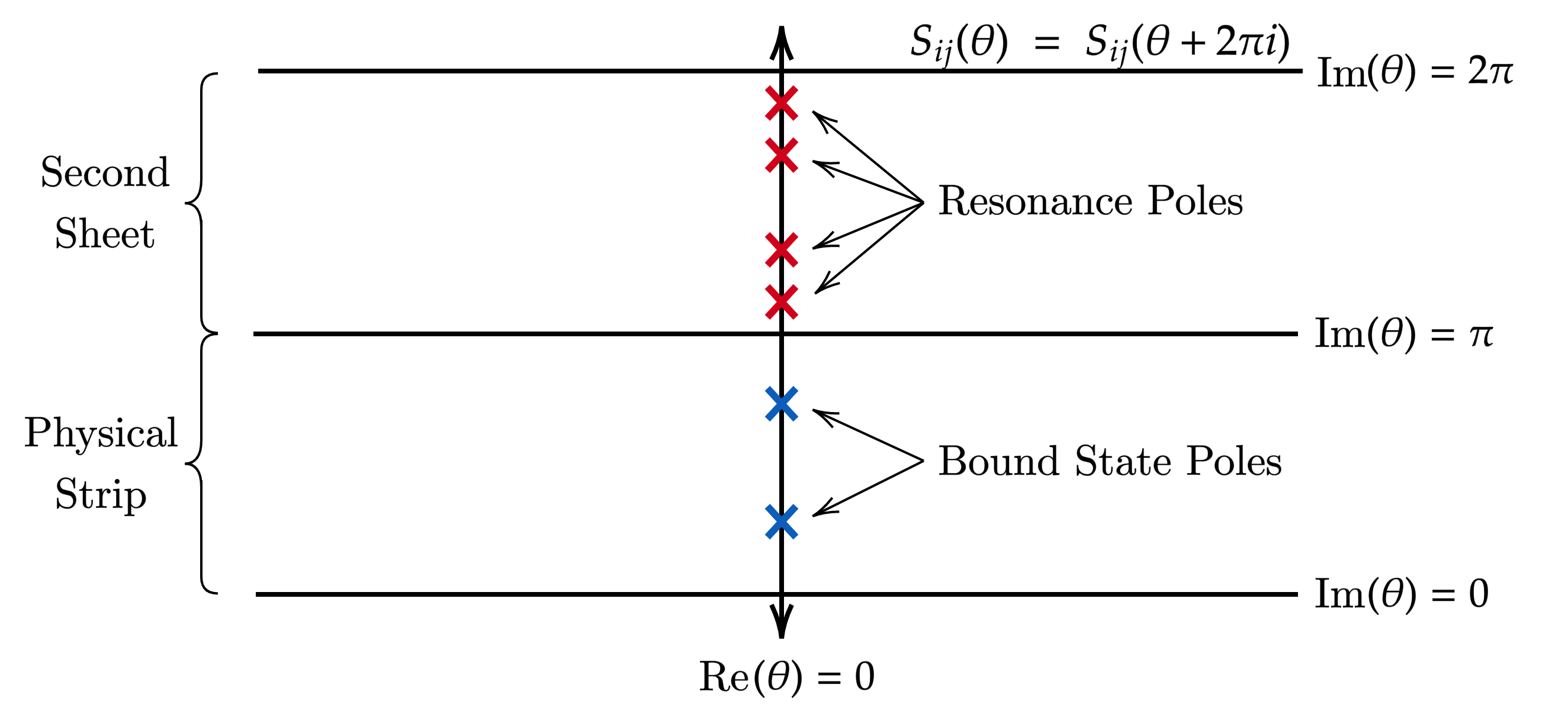}
    \caption{Analytic structure of the $2 \rightarrow 2$ rapidity-space $S$-matrix for purely elastic theories. $S(\theta)$ is periodic with period $2\pi i$, bound state poles appear on the physical strip ${0 \leq \text{Im}(\theta) \leq \pi}$ while resonance poles appear on the second sheet $\pi \leq \text{Im}(\theta) \leq 2\pi$.}
    \label{fig:Structure of 2->2 S-Matrix}
\end{figure}

In what follows, it will be useful to introduce an auxiliary parameter $\widehat{\alpha}_{ij}$ and the function $g_{ij}(\theta)$
\begin{equation}
    \widehat{\alpha}_{ij} = \sum_{\alpha \,\in\, \Omega_{ij}} \hspace{-4pt} \alpha \; , \qquad \qquad g_{ij}(\theta) = \frac{\cosh(\theta + i \pi \widehat{\alpha}_{ij})}{\cosh(\theta)} \;.
\end{equation}
The function $g_{ij}(\theta)$ captures the large-rapidity asymptotics  of the $S$-matrix \eqref{eqn: S-matrix expression} in the sense that
\begin{equation}\label{eq:gij}
    \lim_{\text{Re}(\theta) \rightarrow \pm \infty}[S_{ij}(\theta) - g_{ij}(\theta)] = 0 \; .
\end{equation}
Using linearity of the Fourier transform, the $2 \to 2$ celestial amplitude may be written
\begin{equation}
    \widetilde{M}_{ij}(\Delta) = \int_{-\infty}^\infty d \theta ~ e^{-i \Delta \theta} S_{ij}(\theta) = \int_{-\infty}^\infty e^{-i \Delta \theta} g_{ij}(\theta) + \int_{-\infty}^\infty d\theta ~ e^{-i \Delta \theta} \big[S_{ij}(\theta) - g_{ij}(\theta)\big] \;, 
\end{equation}
where $\Delta \equiv \widehat{\Delta}_2$. Because $g_{ij}(\theta)$ is a relatively simple function, we can compute the first Fourier transform with standard techniques. By contrast, we evaluate the second Fourier transform via contour integration, choosing a rectangular contour encircling both the physical strip and second sheet with sides pushed out to infinity (figure \ref{fig:Structure of 2->2 S-Matrix}). Due to  \eqref{eq:gij}, the vertical sections of this contour do not contribute to the integral. 

Since $g_{ij}(\theta)$ captures the asymptotic behavior of the rapidity-space $S$-matrix at large $\theta$, we will refer to its Fourier transform as the \textit{asymptote term} $A_{ij}(\Delta)$. By contrast, the second Fourier transform is highly theory-dependent and can be expressed in terms of the residues of the poles of $S_{ij}(\theta)$; as such, we refer to it as the \textit{pole term} $P_{ij}(\Delta)$. Altogether, the $2 \to 2$ celestial amplitude is just the sum of these two terms
\begin{equation}
    \widetilde{M}_{ij}(\Delta) = A_{ij}(\Delta) + P_{ij}(\Delta)
    \label{eqn: 2 to 2 celestial amplitude}
\end{equation}
which are given by the following explicit expressions
\begin{equation}\label{eqn: asymptote and pole}
    \begin{split}
        A_{ij}(\Delta) & = 2\pi \cos(\pi \widehat{\alpha}_{ij})\delta(\Delta) + \pi \sin(\pi \widehat{\alpha}_{ij}) \text{csch}\bigg(\frac{\pi \Delta}{2}\bigg) \; ,\\
        P_{ij}(\Delta) &= \frac{2\pi i}{1-e^{2\pi \Delta}}\bigg[ \sum_{\alpha \, \in \, \Omega_{ij}} \text{Res}\Big[e^{- i\Delta \theta}S_{ij}(\theta); \theta = i \pi \alpha\Big] - 2i e^{\Delta \pi} \sin(\pi \widehat{\alpha}_{ij}) \text{cosh}\bigg(\frac{\pi\Delta }{2}\bigg)\bigg] \; .
    \end{split}
\end{equation}
Note that the second term in $P_{ij}(\Delta)$ cancels that of $A_{ij}(\Delta)$ in the final expression. Equations \eqref{eqn: 2 to 2 celestial amplitude} and \eqref{eqn: asymptote and pole} are very useful since they apply to all theories with diagonal scattering and reduce the Fourier transform computation to simple algebra. 

\subsection{Some Examples: Perturbed Diagonal Cosets and Affine Toda Theories}
\label{sec: examples}

In this section, we will introduce two explicit examples of purely elastic $S$-matrices whose properties are determined by a simply laced Lie algebra, $\frak{g}$. These are the perturbed $\widehat{\frak{g}}_1 \oplus \widehat{\frak{g}}_1/\widehat{\frak{g}}_2$ coset CFT and the $\frak{g}$ Affine Toda field theory. In both cases the root system of the algebra informs the set $\Omega_{ij}$ from which the amplitude is constructed. We will not describe the derivation of these $S$-matrices here, and instead refer interested readers to the original papers \cite{Dorey:1990xa,Dorey:1991zp,Braden:1989bu,Fateev:1990hy,Zamolodchikov:1989hfa}. While we focus on these two theories for simplicity, one can use equation \eqref{eqn: 2 to 2 celestial amplitude} to study the celestial $S$-matrices of arbitrary theories characterized by purely elastic scattering.

The coset models have no tunable couplings, but the Affine Toda theories have a continuous coupling constant $\beta \in (0,\infty)$, and their $S$-matrices depend non-trivially on the parameter ${B(\beta) = 2\beta^2/(\beta^2 + 4\pi)}$. They also exhibit S-duality under which  $\beta \mapsto 4\pi/\beta$ (i.e., $B \mapsto 2-B$).

\begin{figure}[H]
    \centering
    \includegraphics[width = .95\textwidth]{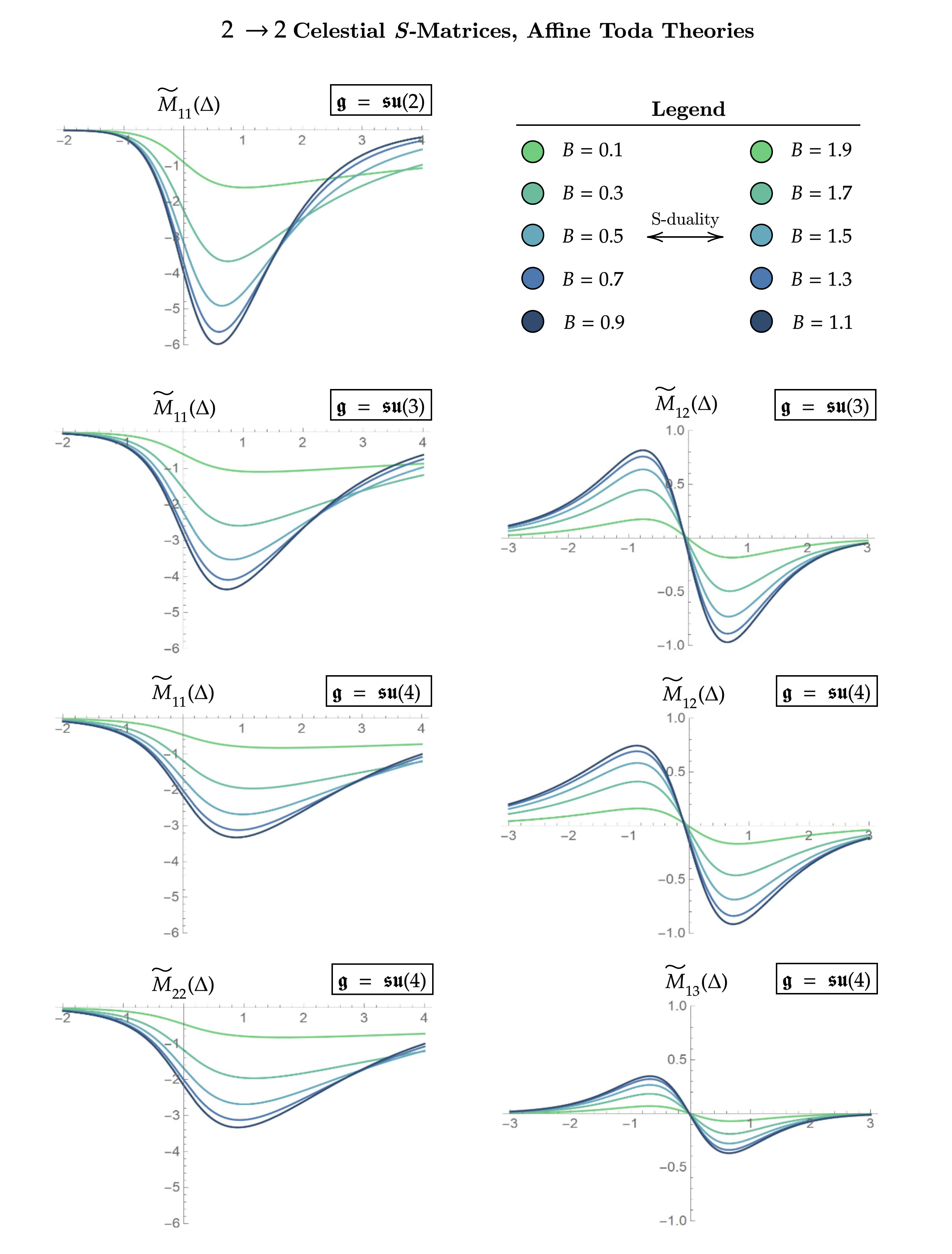}
    \caption{Plots of the Celestial $S$-matrices for the $\frak{g} = \frak{su}(2),\, \frak{su}(3)$, and $\frak{su}(4)$ affine Toda theories suppressing the $A_{ij}(\Delta) = 2\pi \delta(\Delta)$ term. The celestial $S$-matrices satisfy ${\widetilde{M}_{ij}(\Delta) = \widetilde{M}_{ji}(\Delta)}$; moreover, $\widetilde{M}_{11}(\Delta) = \widetilde{M}_{22}(\Delta)$ for the $\frak{su}(3)$ theory while $\widetilde{M}_{11}(\Delta) = \widetilde{M}_{33}(\Delta)$ and $\widetilde{M}_{12}(\Delta) = \widetilde{M}_{32}(\Delta)$ for the $\frak{su}(4)$ theory just as in the case of the deformed coset model.}
    \label{fig: Toda}
\end{figure}

For $\frak{g} = \frak{su}(n)$, both models have particles labelled $j = 1,..., n-1$ with masses $m_j = 2M \sin(\pi j/n)$. The 2-particle rapidity space $S$-matrices satisfy $S_{ij}(\theta) = S_{ji}(\theta)$ and are given by
\begin{equation}
    S_{ij}(\theta) = \prod_{p = 1}^{\text{min}(i,j)}\{i+j+1-2p\}_n
\end{equation}
where the unitary building blocks, $\{x\}_n$, are defined according to
\begin{equation}
    \{x\}_n = \begin{cases} f_{(x-1)/n}(\theta) f_{(x+1)/n}(\theta) \hspace{181pt} \text{Perturbed Coset} \\
    f_{(x-1)/n}(\theta) f_{(x+1)/n}(\theta) f_{(-x+B-1)/n}(\theta) f_{(-x-B+1)/n}(\theta) \hspace{40pt} \text{Affine Toda} \end{cases}
\end{equation}
In table \ref{fig: perturbed coset} we display the celestial amplitudes for the perturbed coset models with $\frak{g} = \frak{su}(2),\frak{su}(3)$, and $\frak{su}(4)$. In figure \ref{fig: Toda}, we plot the celestial amplitude for the Affine Toda theories as a function of $\Delta$ for various choices of the coupling parameter $B$. While it is possible to derive explicit closed form expressions for $\widetilde{M}_{ij}(\Delta)$ in the Toda theories, the results are complicated trigonometric functions of $B$ whose explicit form is unenlightening. Notice that in both cases, these expressions decay exponentially in $\Delta$ as $\Delta \rightarrow \pm \infty$.  We will demonstrate that this is a general feature of CCFT$_0$ correlators in section \ref{sec: properties}.

\begin{table}[h!]
\centering
\begin{tabular}{||c|l|l||} 
\hhline{|t:===:t|}
\textbf{Lie Algebra}     & \multicolumn{1}{c|}{\textbf{$S$-matrix}}  & \multicolumn{1}{c||}{\textbf{Celestial Amplitude}}  \\ 
\hhline{|t:===:t|}
$\frak{su}(2)$ &  $S_{11}(\theta) =  \{1\}_2$     &    $\widetilde{M}_{11}(\Delta) =  -2\pi \delta(\Delta)$                                       \\ 
\hline
$\frak{su}(3)$ &  $S_{11}(\theta) =  \{1\}_3$      &     $\widetilde{M}_{11}(\Delta) = \frac{-2\sqrt{3}\pi}{1-e^{2\pi \Delta}}e^{2\pi \Delta/3} - \pi \delta(\Delta)$                                       \\
      &   $S_{12}(\theta) = \{2\}_3$   &  $\widetilde{M}_{12}(\Delta) = \frac{2\sqrt{3}\pi}{1-e^{2\pi \Delta}} e^{\pi \Delta/3} - \pi \delta(\Delta)$                                          \\ 
\hline
$\frak{su}(4)$ &  $S_{11}(\theta) = \{1\}_4$     & $\widetilde{M}_{11}(\Delta) =  \frac{-4\pi}{1-e^{2\pi \Delta}} e^{\pi \Delta/2}$                                           \\
      &  $S_{22}(\theta) = \{1\}_4\{3\}_4$       &  $\widetilde{M}_{22}(\Delta) =  \frac{8\pi \Delta}{1-e^{2\pi \Delta}} e^{\pi \Delta/2} + 2\pi \delta(\Delta)$                                          \\
      & $S_{12}(\theta) =  \{2\}_4$       & $\widetilde{M}_{12}(\Delta) =  \frac{4\pi}{1-e^{2\pi \Delta}} \Big(e^{\pi \Delta/4}-e^{3\pi \Delta/4}\Big) -2 \pi \delta(\Delta)$                                           \\
      & $S_{13}(\theta) = \{3\}_4$      & $\widetilde{M}_{13}(\Delta) =  \frac{4\pi}{1-e^{2\pi \Delta}} e^{\pi \Delta/2}$                                           \\
\hhline{|t:===:t|}
\end{tabular}
\caption{Celestial amplitudes for the perturbed coset theories with $\frak{g} = \frak{su}(2),\frak{su}(3)$, and $\frak{su}(4)$. There is an additional symmetry exchanging particles $1 \leftrightarrow 2$ in the $\frak{su}(3)$ theory and $1 \leftrightarrow 3$ in the $\frak{su}(4)$ theory relating the amplitudes in the above table to the other possibilities.}
\label{fig: perturbed coset}
\end{table}

Another important feature of the expressions in table \ref{fig: perturbed coset} is the infinite set of poles in the complex $\Delta$ plane located at $\Delta=in$ for $n\in \mathbb{N}$ (equivalently, $\hat{\Delta}\in \mathbb{N}$). This pole structure has also appeared in higher-dimensional examples \cite{Arkani-Hamed:2020gyp}, where it was related to the low-energy and high-energy expansions of the scattering amplitudes. 

One may also use the techniques developed in section \ref{sec:2to2celestialAmp} to compute higher point celestial amplitudes. For example, the $3 \to 3$ celestial amplitudes for the perturbed coset model with $\frak{g} = \frak{su}(3)$ are given by\footnote{There is an additional symmetry swapping particles $1 \leftrightarrow 2$ which relates the above $3 \to 3$ celestial amplitudes to the other four possibilities.}
\begin{align}\label{eqn: 3->3 scattering}
        \widetilde{M}_{111}(\widehat{\Delta}_2,\widehat{\Delta}_3) &= - \frac{2\sqrt{3} \pi}{1-e^{2\pi \widehat{\Delta}_2}} \frac{2\sqrt{3} \pi}{1-e^{2\pi \widehat{\Delta}_3}} e^{\pi (2\widehat{\Delta}_2 + 2\widehat{\Delta}_3)/3} ~ ~ + ~ \cdots \notag\\
        \widetilde{M}_{112}(\widehat{\Delta}_2,\widehat{\Delta}_3) &= \frac{2\sqrt{3} \pi}{1-e^{2\pi \widehat{\Delta}_3}} \frac{2\sqrt{3} \pi}{1-e^{2\pi (\widehat{\Delta}_2 + \widehat{\Delta}_3)}} e^{\pi (2\widehat{\Delta}_2 + 3\widehat{\Delta}_3)/3} ~ ~ + ~ \cdots \\
        \widetilde{M}_{121}(\widehat{\Delta}_2,\widehat{\Delta}_3) &= \frac{2\sqrt{3}\pi}{1-e^{2\pi \widehat{\Delta}_3}}\frac{6\pi \widehat{\Delta}_3}{1-e^{2\pi (\widehat{\Delta}_2 + \widehat{\Delta}_3)}} e^{\pi(\widehat{\Delta}_2 + 2 \widehat{\Delta}_3)/3} + \cdots  \notag\\
        \widetilde{M}_{211}(\widehat{\Delta}_2,\widehat{\Delta}_3) &= - \frac{2\sqrt{3} \pi}{1-e^{2\pi \widehat{\Delta}_2}}\frac{2\sqrt{3} \pi}{1-e^{2\pi \widehat{\Delta}_3}} e^{\pi(\widehat{\Delta}_2 + \widehat{\Delta}_3)/3} + \frac{2\sqrt{3} \pi}{1-e^{2\pi \widehat{\Delta}_3}}\frac{2\sqrt{3} \pi}{1-e^{2\pi (\widehat{\Delta}_2 + \widehat{\Delta}_3)}} e^{\pi(\widehat{\Delta}_2 + 3 \widehat{\Delta}_3)/3} + \cdots \notag
\end{align}
where we have suppressed terms which have $\delta(\widehat{\Delta}_i)$ singularities in the `$\cdots$' terms. The full expression is given in Appendix \ref{appendix: n to n amplitudes}, where we explain how to compute these amplitudes and outline the inductive procedure to compute $n \to n$ celestial amplitudes for all integrable theories with diagonal scattering.

\subsection{Properties of the $2$-Particle Celestial Amplitude}
\label{sec: properties}

In this section, we will study the behavior of the celestial amplitude at large and small $\Delta$.

\subsubsection*{Exponential Decay of the Celestial Amplitude}

In the previous section we noted that the celestial amplitude decays exponentially as $\Delta \rightarrow \pm \infty$ for the affine Toda and perturbed coset models. This behavior actually holds for all models with diagonal scattering and follows directly from \eqref{eqn: asymptote and pole}. First, note that
\begin{equation}
    \widetilde{M}_{ij}(\Delta) \sim 2\pi i \, e^{-\pi (\Delta + |\Delta|)} \sum_{\alpha \, \in \, \Omega_{ij}} \text{Res}\Big[e^{-i \Delta \theta} S_{ij}(\theta); \theta = i \pi \alpha\Big] \quad \text{as} \quad \Delta \rightarrow \pm \infty \; .
\end{equation}
When $S_{ij}(\theta)$ has only simple poles, the residue calculation is straightforward
\begin{equation}
    \widetilde{M}_{ij}(\Delta) \sim 2\pi i \sum_{\alpha \, \in \, \Omega_{ij}} \text{Res}[S_{ij}(\theta); \theta = i \pi \alpha] \times \begin{cases} e^{-\pi (2-\alpha) \Delta} \hspace{30pt} \Delta \rightarrow + \infty\\ e^{\pi \alpha \Delta} \hspace{54pt} \Delta \rightarrow - \infty \end{cases}
    \label{eqn: decay of non-dressed}
\end{equation}
Since $\alpha \in (0,2)$, the celestial amplitude is well approximated at large $|\Delta|$ by a finite sum of terms, each of which is exponentially suppressed. The celestial amplitude therefore decays exponentially in $|\Delta|$, and the rate of decay is fixed by the parameter $\alpha$ which encodes the positions of the poles (i.e., the dynamics of the model). Bound state poles are typically not visible at any order in perturbation theory, so we conclude that the non-perturbative behavior of the model controls crucial aspects of the analytic structure of the celestial correlators.
Indeed, note that when $\theta = i \pi \alpha$ is a bound state pole corresponding to forward-channel scattering, the value of $\alpha$ determines the mass of the bound state according to equation \eqref{eqn: bound state mass}. Moreover, the residue $ \text{Res}[S_{ij}(\theta); \theta = i \pi \alpha]$ is related to the three-point coupling between the two external particles and the bound state, $f_{ijB}$, according to equation \eqref{eqn: bound state coupling}. Therefore, the rate of exponential decay of the celestial amplitude encodes highly non-trivial information about the spectrum and coupling constants of the underlying QFT. It would be interesting to establish this relationship in non-integrable models, where it could provide a non-perturbative axiom for the definition of CCFT. 

This discussion assumed that all poles of $S_{ij}(\theta)$ are simple, but the conclusions still apply when the poles are of higher order. The only dependence of the residue on $\Delta$ enters through the $e^{-i\Delta \theta}$ term. All others factors are bounded functions of $\alpha$ which do not affect the asymptotics. However, the interpretation of the higher-order poles is not as direct in this case (they are related to anomalous threshold singularities, which produce poles rather than cuts in two dimensions).

\subsubsection*{Unboundedness of the Celestial Amplitude}

Equation \eqref{eqn: asymptote and pole} indicates that the celestial amplitude is ill-defined as a function when $\Delta \rightarrow 0$. Indeed at $\Delta = 0$, the asymptote term has a delta function singularity along with a term that blows up as $\tfrac{1}{\Delta}$. Both behaviors have simple explanations. The $\delta(\Delta)$ term arises because the two-particle $S$-matrix is not an integrable function, and the zero mode of the Fourier transform is simply the integral of $S_{ij}(\theta)$, which diverges. 
The $\tfrac{1}{\Delta}$ behavior is a reflection of the fact that the $\theta \rightarrow +\infty$ limit of $S_{ij}(\theta)$ differs from the $\theta \rightarrow - \infty$ limit.  For small $\Delta$ (i.e., long wavelength), $g_{ij}(\theta)$ looks like a step function whose Fourier transform has a $\tfrac{1}{\Delta}$ pole. In non-integrable models, the large-rapidity behavior of $S_{ij}(\theta)$ is dampened due to particle production, and this could potentially soften the singular behavior of the Fourier transform. Unfortunately, there is little rigorous information available for non-integrable $S$-matrices in the large-rapidity regime, and it is an important open question to determine whether the singular behavior we observe here persists in non-integrable models.

In contrast, the pole term $P_{ij}(\Delta)$ is finite as $\Delta \rightarrow 0$ despite the $(1-e^{2\pi \Delta})^{-1}$ prefactor. Indeed, the expression inside of the parentheses actually vanishes as $\Delta \rightarrow 0$, so the product has a finite limit. The fact that $P_{ij}(\Delta)$ is a bounded function follows from the fact that it is the Fourier transform of $S_{ij}(\theta) - g_{ij}(\theta)$. One can verify that $S_{ij}(\theta) - g_{ij}(\theta)$ is a Schwartz class function, so its Fourier transform will be bounded.

The distributional nature of the celestial amplitude appears to be a generic feature of local quantum field theory. However, as in AdS holography, we do not expect celestial CFTs to be fully well-behaved without the inclusion of gravity. Of course, two-dimensional gravity lacks many of the important features of higher-dimensional gravity, so the expectation is less clear in our case. Nevertheless, in section \ref{sec: dressing boundedness} we will see that coupling these models to even the simplest version of 2d gravity dramatically changes the analytic properties of the celestial correlators. No matter how strong or weak the coupling to gravity is, the full celestial amplitude immediately becomes a bounded function. Gravity tames the divergences and smooths out the bad behavior.

\section{CCFT Dual to Integrable Theories III: Gravitational Dressing}
\label{sec:grav}

Since gravity has no propagating degrees of freedom in two dimensions, its signature in the $S$-matrix is slightly more subtle than in higher dimensions. Nonetheless, although there are no gravitons to propagate in loops or appear in external states, the path integral over metrics can still have interesting implications for observables. For instance, one generally expects that local observables will cease to be sharply defined upon coupling to gravity. Boundary observables like the $S$-matrix are still believed to make sense, and it is generally hoped that they can be computed holographically. 

Given that 2d quantum gravity really just means integration over the conformal mode plus some boundary degrees of freedom, one does not expect coupling to gravity to dramatically alter the spectrum of a QFT in flat space. At most one has an extra Liouville-like degree of freedom to integrate over, but the simplest theories of gravity in 2d are quasi-topological and do not even include that mode. Gravity is most often introduced through the Lagrangian, but the authors of \cite{Dubovsky:2013ira} discovered a general procedure for ``dressing'' non-gravitational 2d $S$-matrices that exhibits many gravitational traits. The deformation preserves the spectrum of particles (as well as integrability when it is present) but significantly changes the UV behavior of amplitudes. Importantly, this dressing procedure is equivalent to coupling the model to flat space JT gravity \cite{Dubovsky:2017cnj}.

The dressing procedure takes a simple form for the $2\to 2$ $S$-matrix of an integrable model. One simply introduces a UV scale $l_s$ and dresses the 2-particle amplitude by a rapidity-dependent phase
\begin{equation}\label{eq:dressed}
    S_{a_1 a_2}^{b_1 b_2}(\theta)_{\text{dressed}} = e^{il_s^2 m_i m_j \sinh(\theta)} S_{a_1 a_2}^{b_1 b_2}(\theta) \; .
\end{equation}
The dressed $n \to n$ rapidity-space amplitude remains the product of $2 \to 2$ scattering amplitudes; however, these constituent amplitudes are now all individually dressed. One may verify that these dressed $S$-matrices also obey real-analyticity, unitarity, and crossing symmetry, so the corresponding constraints on celestial amplitudes are unmodified in the presence of gravity. Indeed, phases like \eqref{eq:dressed} satisfy all the properties of local QFT except for polynomial boundedness: the dressing has an essential singularity at $s=(p_1+p_2)^2= \infty$ which distinguishes it from local QFT and which is responsible for the non-local properties of the deformed model. 

To simplify some technical aspects of the discussion, we will focus on diagonal scattering. Nevertheless, the dressing procedure described in \cite{Dubovsky:2017cnj} can be applied to any two-dimensional QFT and many of the following results probably hold more generally. We hope to explore this in future work.

\subsection{The Celestial Dressing}\label{sec: celestial dressing}
The dressed $2 \to 2$ celestial amplitude is given by the Fourier transform
\begin{equation}
    \widetilde{M}_{ij}(\Delta)_{\text{dressed}} = \int_{-\infty}^\infty d\theta ~ e^{-i \Delta \theta} e^{il_s^2 m_i m_j \sinh(\theta)}S_{ij}(\theta) \; .
\end{equation}
It can be computed by convolving the undressed celestial amplitude $\widetilde{M}_{ij}(\Delta)$ with the \textit{celestial dressing} $\widetilde{D}_{ij}(\Delta)$
\begin{equation}\label{eqn: convolution}
    \widetilde{M}_{ij}(\Delta)_{\text{dressed}} = \int_{-\infty}^\infty d \Delta' ~ \widetilde{D}_{ij} (\Delta') \widetilde{M}_{ij}(\Delta - \Delta') \;,
\end{equation}
where $\widetilde{D}_{ij}(\Delta)$ is the Fourier transform of the gravitational dressing
\begin{equation}    \label{eqn: def celestial dressing}
    \widetilde{D}_{ij}(\Delta) = \int_{-\infty}^\infty d \theta ~  e^{-i \Delta \theta} e^{il_s^2 m_i m_j \sinh(\theta)} \; .
\end{equation}
We have already discussed the Fourier transform of the undressed amplitude extensively, so it remains to compute this celestial dressing. The representation
\begin{equation}
    \widetilde{D}_{ij}(\Delta) = \int_{-\infty}^\infty d \theta ~ \cos\big(\Delta \theta - l_s^2 m_i m_j \sinh(\theta)\big)
\end{equation}
makes clear that $\text{Im} \; \widetilde{D}_{ij}(\Delta) = 0$ as is necessary to be consistent with real analyticity. 

\begin{figure}
    \centering
    \includegraphics[width = \textwidth]{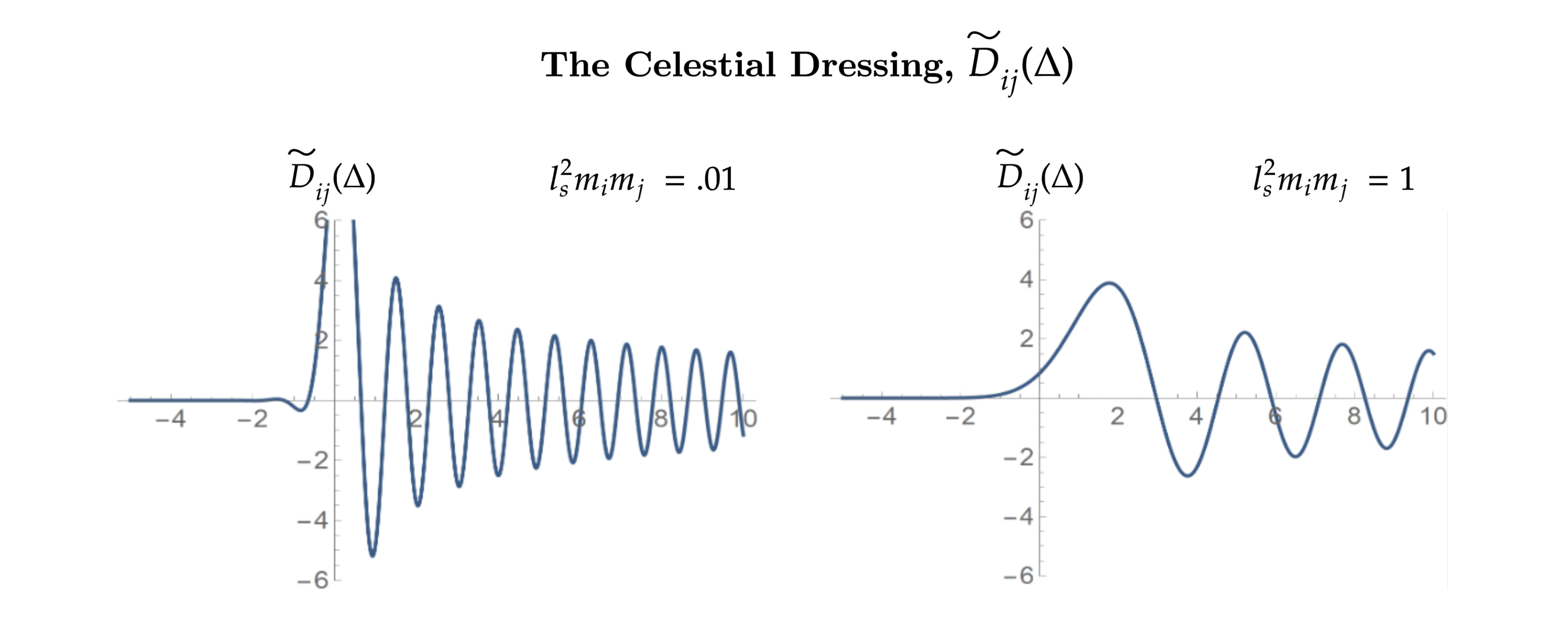}
    \caption{The celestial dressing, $\widetilde{D}_{ij}(\Delta)$, for $l^2_s m_i m_j = 0.01$ (weak gravity, left hand panel), for $l^2_s m_i m_j = 1$ (stronger gravity, right hand panel).}
    \label{fig: dressing}
\end{figure}

Perhaps surprisingly, it is possible to do the integral \eqref{eqn: def celestial dressing} exactly.
The Bessel function of the first kind has an integral representation
\begin{equation}
    J_\nu(x)= \frac{1}{2\pi i}\int_{\infty -i\pi}^{\infty + i\pi} e^{x\sinh \theta -\nu \theta} d\theta \; ,
\end{equation}
from which it follows that
\begin{equation}
    J_{-\nu}(-x)=\frac{1}{2\pi i}\int_{-\infty -i\pi}^{-\infty +i\pi}e^{x\sinh \theta -\nu \theta} d\theta \; .
\end{equation}
To evaluate \eqref{eqn: def celestial dressing}, we consider the semi-infinite rectangular contour in the complex $\theta$ plane with $\text{Im}(\theta)\in [-\pi,\pi]$ which encloses no poles:
\begin{equation}
    \int_{-\infty-i\pi}^{\infty-i\pi} e^{-x\sinh \theta -\nu \theta}d\theta
    + \int_{\infty-i\pi}^{\infty+i\pi} e^{-x\sinh \theta -\nu \theta}d\theta
    - \int_{-\infty+i\pi}^{\infty+i\pi} e^{-x\sinh \theta -\nu \theta}d\theta
    -\int_{-\infty-i\pi}^{-\infty+i\pi} e^{-x\sinh \theta -\nu \theta}d\theta = 0 \; .
\end{equation}
From this one concludes that 
\begin{equation}
    (e^{i \pi \nu} - e^{-i\pi \nu})\int_{-\infty}^{\infty} e^{x\sinh \theta -\nu \theta}d\theta
    =2\pi i \left[J_{-\nu}(x)-J_\nu(-x)\right] \; .
\end{equation}
The celestial dressing is therefore a sum of Bessel functions
\begin{equation}\label{eq:dressingFT}
    \widetilde{D}_{ij}(\Delta)=i\pi\frac{\left[J_{i\Delta}(-il^2_s m_i m_j)- J_{-i\Delta}(il^2_s m_i m_j)\right]}{\sinh \pi \Delta} \; .
\end{equation}
We plot this function for some representative values of $l_s^2m_im_j$ in figure \ref{fig: dressing}. One can also numerically integrate \eqref{eqn: def celestial dressing} and verify that it matches \eqref{eq:dressingFT}.

\subsection{Boundedness of the Dressed Celestial Amplitude}
\label{sec: dressing boundedness}

We saw in  section \ref{sec: properties} that the $n \to n$ celestial amplitude generally only exists as a tempered distribution in the absence of gravity. In particular, the celestial amplitude is generically unbounded as $\Delta \rightarrow 0$. 
These results all assumed the usual axioms of local quantum field theory, in particular polynomial boundedness of the amplitudes in the Mandelstam variables. The rapidly oscillating phase \eqref{eq:dressed} has the potential to dramatically effect these conclusions, and in this section we discuss to what extent these features are modified by gravity.

Rapidly varying phases in the Fourier transform often lead to better convergence properties. We would like to know if there exists a constant $\widetilde{M}_{\text{max}}< \infty$ such that
\begin{equation}
    |\widetilde{M}_{ij}(\Delta)| \leq \widetilde{M}_{\text{max}} \qquad \qquad \forall ~ \Delta \in \mathbb{R} \; .
\end{equation}
This would demonstrate that the dressed celestial amplitude is a bounded function that doesn't display the singular distributional features of the undressed theory. We begin by demonstrating that the following integral converges
\begin{equation}
    \begin{split}
        \widetilde{M}_{ij}(0)_{\text{dressed}} &= \int_{-\infty}^\infty d\theta ~ e^{i l_s^2 m_i m_j \sinh(\theta)} S_{ij}(\theta) \\
        &= \frac{1}{il_s^2 m_i m_j} \int_{-\infty}^\infty d\theta ~ \frac{1}{\cosh(\theta)} \frac{d}{d\theta}\bigg[ e^{i l_s^2 m_i m_j \sinh(\theta)}\bigg] S_{ij}(\theta) \\
        &= -\frac{1}{il_s^2 m_im_j} \int_{-\infty}^\infty d\theta ~ e^{il_s^2 m_i m_j \sinh(\theta)} \bigg[\frac{S_{ij}'(\theta)}{\cosh(\theta)} - \frac{S_{ij}(\theta) \sinh(\theta)}{\cosh(\theta)^2}\bigg] < \infty
    \end{split}
\end{equation}
where we integrated by parts and dropped the boundary term due to the rapid decay of $1/\cosh(\theta)$ as $\theta \rightarrow \pm \infty$. In fact the same would hold if we restored the $e^{-i\Delta \theta}$ term.  Indeed, if we insert the explicit expression for $S_{ij}(\theta)$ from equation \eqref{eqn: S-matrix expression}, we see that the integral is absolutely convergent since the modulus of the integrand is exponentially decaying as $\theta \rightarrow \pm \infty$. It is known that the Fourier transform of an integrable function is a uniformly continuous function and is bounded by the value of the absolutely converging integral, $\widetilde{M}_{\text{max}}$. It follows that $\widetilde{M}_{ij}(\Delta)_{\text{dressed}}$ is a bounded function.

\section{Discussion}\label{sec:Discussion}
This paper contains some of the first exact results in the celestial CFT program, but we have only scratched the surface of the set of interesting models in two dimensions. The most straightforward generalization would be to study the 2d integrable $S$-matrices with non-diagonal scattering. This class contains a variety of important quantum field theories, including the nonlinear sigma models, the Gross-Neveu model, and the Sinh-Gordon model. It would be considerably more interesting to study non-integrable models which nonetheless admit some degree of analytic control, perhaps by utilizing the large-$N$ expansion. The t'Hooft model seems like a promising target in this regard. More generally, two-dimensional gauge theory lacks propagating degrees of freedom but does produce important global effects that would be interesting to explore in CCFT$_0$. 
Theories with massless particles present additional challenges due to the strong infrared divergences in low dimensions, but results do exist \cite{Fendley:1993jh} that could provide new insight into CCFT.

The characterization of the asymptotics of the celestial correlators in terms of the bound state spectrum is the type of non-perturbative structural result we seek in higher dimensions. In the presence of inelastic scattering, multiparticle cuts are expected and the analytic structure of the $2\to2$ amplitude is much more complicated. Nonetheless, results in this broader class of tractable models might provide important structural clues for the higher dimensional case.

The most interesting aspect of this work is the dramatic effect of the gravitational dressing. Many models of 2d quantum gravity can be solved exactly using matrix model techniques and would be interesting to explore within CCFT$_0$, but even the simplest model of 2d gravity manages to smooth out the distributional nature of the celestial correlators. It would be fascinating if this mechanism also operates in higher dimensions. Indeed, the phase shift $e^{il_s^2 s}$ also occurs in higher-dimensional high-energy elastic scattering, where it can be derived by summing ladder diagrams or performing a geodesic calculation using shocks \cite{Dray:1984ha}. In these cases $l_s^2\sim G_Nb^{d-2}$ is related to the impact parameter and there is dependence on the transverse coordinates, but it seems likely that the basic mechanism exhibited in this paper will also smooth out higher-dimensional celestial correlators (see \cite{PipolodeGioia:2022exe} for work in this direction). These phase shifts and time delays are also related to chaos in the black hole $S$-matrix \cite{Shenker:2013pqa,Polchinski:2015cea} and it would be interesting to explore this connection within the CCFT context.

\section*{Acknowledgements}

We would like to thank Barak Gabai, J. Alexander Jacoby, Puskar Mondal, Andrew Strominger, and Xi Yin for stimulating discussions. We would also like to thank Monica Pate and Ana-Maria Raclariu for useful comments on the draft. The work of DK is supported by the Harvard Center of Mathematical Sciences and Applications and DOE grant de-sc/0007870. AT gratefully acknowledges support from NSF GRFP grant
DGE1745303.

\appendix
\section{Implications of Crossing Symmetry}
\label{appendix:crossing}

In this appendix, we will derive the celestial analog of the crossing equation 
\begin{equation}
    S_{a_1 a_2}^{b_1 b_2}(\theta)_{\text{dressed}} = S_{a_1 \mathcal{C}(b_2)}^{b_1 \mathcal{C}(a_2)}(i \pi-\theta)_{\text{dressed}} \; ,
\end{equation} 
where $\mathcal{C}$ is the charge conjugation operator. The result holds for both integrable and non-integrable quantum field theories and for any choice of the gravitational dressing $l_s^2 > 0$.

To compute this Fourier transform we use the contour integration techniques of section \ref{sec:2to2celestialAmp}. For this calculation we choose a contour $\gamma$ surrounding the physical strip but not the second sheet
\begin{align}
        \int_\gamma &d\theta ~ e^{- i \Delta \theta} S_{a_1 a_2}^{b_1 b_2}(\theta)_{\text{dressed}} = \bigg(\int_{-\infty}^\infty   + \int_{\infty}^{\infty + i \pi}  + \int_{\infty + i \pi}^{-\infty + i \pi}  + \int_{-\infty + i \pi}^{-\infty} \bigg) ~ e^{-i\Delta \theta} S_{a_1 a_2}^{b_1 b_2}(\theta)_{\text{dressed}} \\
        &= \widetilde{M}_{a_1 a_2}^{b_1 b_2}(\Delta)_{\text{dressed}} - e^{\pi \Delta} \widetilde{M}_{a_1 \mathcal{C}(b_2)}^{b_1 \mathcal{C}(a_2)} (-\Delta)_{\text{dressed}} 
      + \bigg(\int_{\infty}^{\infty + i \pi} + \int_{-\infty + i \pi}^{-\infty} \bigg) e^{-i\Delta \theta} S_{a_1 a_2}^{b_1 b_2}(\theta)_{\text{dressed}} \; .\notag
\end{align}
Next, we demonstrate that the remaining vertical sections of the contour vanish. Decomposing $\theta = x + iy$, the integrand on the right side of the contour takes the form
\begin{equation}
    \lim_{x \rightarrow \infty} \int_0^{\pi} d y ~ e^{-i\Delta (x+iy)} S_{a_1 a_2}^{b_1 b_2}(x+iy) e^{il_s^2 \sinh(x + iy)} \; .
\end{equation}
To show that the integral vanishes, we bound the modulus of the three terms in the product separately for both $x \rightarrow \infty$ and $x \rightarrow -\infty$
\begin{equation}
    \begin{split}
        |e^{-i \Delta(x + i y)}| = e^{\Delta y}\; , \quad
        |S_{a_1 a_2}^{b_1 b_2}(x + i y)| &\leq e^{\kappa |x+y|}\; , \quad
        |e^{il_s^2m_i m_j \sinh(x + iy)}| = e^{-l_s^2 m_i m_j\sin y\cosh x} \; .
    \end{split}
\end{equation}
The second bound follows from the standard assumption in local quantum field theory of polynomial boundedness in the Mandelstam invariant $s$. Since $y \in (0,\pi)$ on the physical strip, the vertical sections of the integral decay superexponentially and have vanishing contribution to the full integral. 

Evaluating the contour integral explicitly with residues gives the final result
\begin{equation}
    \widetilde{M}_{a_1 a_2}^{b_1 b_2}(\Delta)_{\text{dressed}} - e^{\pi \Delta} \widetilde{M}_{a_1 \mathcal{C}(b_2)}^{b_1 \mathcal{C}(a_2)} (-\Delta)_{\text{dressed}} = 2 \pi i \sum_{\alpha \,\in \,\Omega_{BS}} \text{Res}\Big[e^{-i \Delta \theta}S_{a_1 a_2}^{b_1 b_2}(\theta)_{\text{dressed}};\theta = i \pi \alpha\Big] \; ,
\end{equation}
where $\Omega_{BS}$ is the set of bound state poles in the $a_1 + a_2 \rightarrow b_1 + b_2$ scattering amplitude. A similar calculation could be performed in the undressed theory by utilizing the function $g_{ij}(\theta)$ to remove the vertical contributions.

\section{$n \to n$ Celestial Amplitudes}
\label{appendix: n to n amplitudes}

In this appendix we explicitly compute the $3 \to 3$ celestial amplitude for the deformed coset model discussed in section \ref{sec: examples} for $\frak{g} = \frak{su}(3)$ and explain how this approach may be generalized to compute the $n \to n$ celestial amplitude for all integrable models with diagonal scattering.

The $3 \to 3$ celestial amplitude is given by equation \eqref{eqn:NtoNCelestial}
\begin{equation}
    \begin{split}
        \widetilde{M}_{ijk}(\widehat{\Delta}_2,\widehat{\Delta}_3) &= \int_{-\infty}^\infty d\theta_{12} ~ e^{-i \widehat{\Delta}_2 \theta_{12}} S_{ij}(\theta_{12}) \int_{-\infty}^\infty d\theta_{13} ~ e^{-i \widehat{\Delta}_3 \theta_{13}} S_{ik}(\theta_{13}) S_{jk}(\theta_{13}-\theta_{12}) \; ,
    \end{split}
\end{equation}
where we have used factorizability of the rapidity-space $S$-matrix to write the $3 \to 3$ amplitude as a product of $2 \to 2$ amplitudes. The integral over $\theta_{13}$ may be evaluated explicitly using the contour integration arguments of section \ref{sec:2to2celestialAmp}. Concretely, we define
\begin{equation}
    I_{ijk}(\theta_{12},\widehat{\Delta}_3) = \int_{-\infty}^\infty d\theta_{13} ~ e^{-i \widehat{\Delta}_3 \theta_{13}} S_{ik}(\theta_{13}) S_{jk}(\theta_{13}-\theta_{12})
\end{equation}
which may be computed by taking residues of the integrand at $\theta_{13} \in i \pi \Omega_{ik}$ and at $\theta_{13} \in \theta_{12} + i \pi \Omega_{jk}$
\begin{equation}
    \begin{split}
        I_{111} &= \frac{-2\sqrt{3}\pi}{1-e^{2\pi \widehat{\Delta}_3}}\bigg[e^{2\pi \widehat{\Delta}_3/3} \hspace{4pt} \frac{\sinh\big(\frac{1}{2}(\theta_{12}- \frac{4\pi i}{3})\big)}{\sinh\big(\frac{1}{2} \theta_{12}\big)} + e^{-i \widehat{\Delta}_3 \theta_{12}} e^{2\pi \widehat{\Delta}_3/3} \hspace{4pt} \frac{\sinh\big(\frac{1}{2}(\theta_{12}+ \frac{4\pi i}{3})\big)}{\sinh\big(\frac{1}{2} \theta_{12}\big)}\bigg] -\pi \delta(\widehat{\Delta}_3)\; ,\\
        I_{112} &=  \frac{-2\sqrt{3}\pi}{1-e^{2\pi \widehat{\Delta}_3}}\bigg[e^{\pi \widehat{\Delta}_3/3} \hspace{4pt} \frac{\sinh\big(\frac{1}{2}(\theta_{12}- \frac{2\pi i}{3})\big)}{\sinh\big(\frac{1}{2} \theta_{12}\big)} + e^{-i \widehat{\Delta}_3 \theta_{12}} e^{\pi \widehat{\Delta}_3/3} \hspace{4pt} \frac{\sinh\big(\frac{1}{2}(\theta_{12}+ \frac{2\pi i}{3})\big)}{\sinh\big(\frac{1}{2} \theta_{12}\big)}\bigg] -\pi \delta(\widehat{\Delta}_3)\;,\\
        I_{121} &= \frac{-2\sqrt{3}\pi}{1-e^{2\pi \widehat{\Delta}_3}}\bigg[e^{2\pi \widehat{\Delta}_3/3} \hspace{4pt} \frac{\sinh\big(\frac{1}{2}(\theta_{12}+ i\pi)\big)}{\sinh\big(\frac{1}{2}( \theta_{12}-\frac{i\pi}{3})\big)} 
        - e^{-i \widehat{\Delta}_3 \theta_{12}} e^{\pi \widehat{\Delta}_3/3} \hspace{4pt} \frac{\sinh\big(\frac{1}{2}(\theta_{12}+i\pi)\big)}{\sinh\big(\frac{1}{2} (\theta_{12}-\frac{i\pi}{3})\big)}\bigg] +2\pi \delta(\widehat{\Delta}_3)\; ,\\
        I_{211} &= \frac{-2\sqrt{3}\pi}{1-e^{2\pi \widehat{\Delta}_3}}\bigg[e^{\pi \widehat{\Delta}_3/3} \hspace{4pt} \frac{\sinh\big(\frac{1}{2}(\theta_{12}+ i\pi)\big)}{\sinh\big(\frac{1}{2}( \theta_{12}+\frac{i\pi}{3})\big)} 
        - e^{-i \widehat{\Delta}_3 \theta_{12}} e^{2 \pi \widehat{\Delta}_3/3} \hspace{4pt} \frac{\sinh\big(\frac{1}{2}(\theta_{12}+i\pi)\big)}{\sinh\big(\frac{1}{2} (\theta_{12}+\frac{i\pi}{3})\big)}\bigg] +2\pi \delta(\widehat{\Delta}_3) \; .
        \nonumber
    \end{split}
\end{equation}
These integrals are also related by a $1 \leftrightarrow 2$ symmetry inherent in the rapidity space $S$-matrix. These expressions are very similar to the most general $2 \to 2$ rapidity-space scattering amplitude given by equation \eqref{eqn: S-matrix expression}. Thus, one may evaluate the outer integral by performing one more set of straightforward residue computations. The final result is
\begin{equation}
    \begin{split}
        \widetilde{M}_{111} &= -m_2(\widehat{\Delta}_3)m_2(\widehat{\Delta}_2) \\
        &\hspace{30pt} - m_2(\widehat{\Delta}_3) \cdot 2\pi \delta(\widehat{\Delta}_2) + m_2(\widehat{\Delta}_3) \cdot \pi \delta(\widehat{\Delta}_1) + m_2(\widehat{\Delta}_2)\cdot \pi \delta(\widehat{\Delta}_3) +\pi^2 \delta(\widehat{\Delta}_2) \delta(\widehat{\Delta}_3) \; ,\\
        \widetilde{M}_{112} &= -m_1(\widehat{\Delta}_3) m_4(\widehat{\Delta}_1) \\
        &\hspace{30pt} -m_1(\widehat{\Delta}_3) \cdot \pi \delta(\widehat{\Delta}_2) + m_1(\widehat{\Delta}_3) \cdot 2\pi \delta(\widehat{\Delta}_1)+ m_2(\widehat{\Delta}_2) \cdot \pi \delta(\widehat{\Delta}_3) + \pi^2 \delta(\widehat{\Delta}_2)\delta(\widehat{\Delta}_3) \; , \\
        \widetilde{M}_{121} &= - \sqrt{3}\widehat{\Delta}_2 m_2(\widehat{\Delta}_3) m_1(\widehat{\Delta}_2) + \sqrt{3}\widehat{\Delta}_1 m_1(\widehat{\Delta}_3)m_5(\widehat{\Delta}_1) \\
        &\hspace{30pt} - m_2(\widehat{\Delta}_3) \cdot 2 \pi \delta(\widehat{\Delta}_2) + m_1(\widehat{\Delta}_3) \cdot \pi \delta(\widehat{\Delta}_1) + m_1(\widehat{\Delta}_2) \cdot 2\pi \delta(\widehat{\Delta}_3) - 2\pi^2 \delta(\widehat{\Delta}_2) \delta(\widehat{\Delta}_3) \; ,\\
        \widetilde{M}_{211} &= -m_1(\widehat{\Delta}_3)m_1(\widehat{\Delta}_2) -m_5(\widehat{\Delta}_1) m_2(\widehat{\Delta}_3) \\
        &\hspace{30pt} -m_1(\widehat{\Delta}_3) \cdot \pi \delta(\widehat{\Delta}_2) + m_2(\widehat{\Delta}_3) \cdot 2\pi\delta(\widehat{\Delta}_1) +  m_1(\widehat{\Delta}_2) \cdot 2\pi \delta(\widehat{\Delta}_3) - 2\pi^2 \delta(\widehat{\Delta}_2) \delta(\widehat{\Delta}_3) \; ,\\
    \end{split}
\end{equation}
where we have defined the following function to simplify our expressions
\begin{equation}
    m_\alpha(\Delta) = \frac{2\sqrt{3} \pi}{1-e^{2\pi \Delta}} e^{\pi \alpha \Delta/3} \;.
\end{equation}
In each case, the first line only includes terms without a $\delta(\widehat{\Delta}_i)$ singularity (compare with equation \eqref{eqn: 3->3 scattering}), while the second line includes terms which have one or two such delta functions.
We have also set $\widehat{\Delta}_1 = -\widehat{\Delta}_2 - \widehat{\Delta}_3$, which holds on the delta function support of the celestial $S$-matrix as in equation \eqref{eqn:S-matrixdeltaconstraint}. 

Generalizing this approach to $n \to n$ scattering in an arbitrary integrable theory is a straightforward task. One uses factorizability to write the $n \to n$ rapidity space amplitude as a product of $2 \to 2$ amplitudes. The celestial amplitude is an $n-1$ dimensional Fourier transform over the rapidity space variables. These Fourier transforms may be computed successively through repeated residue computations at the poles of the constituent $2 \to 2$ amplitudes.

\bibliographystyle{apsrev4-1long}
\bibliography{bib.bib}

\begin{thebibliography}{10}%
\makeatletter
\providecommand \@ifxundefined [1]{%
 \ifx #1\undefined \expandafter \@firstoftwo
 \else \expandafter \@secondoftwo
\fi
}%
\providecommand \@ifnum [1]{%
 \ifnum #1\expandafter \@firstoftwo
 \else \expandafter \@secondoftwo
\fi
}%
\providecommand \enquote [1]{``#1''}%
\providecommand \bibnamefont  [1]{#1}%
\providecommand \bibfnamefont [1]{#1}%
\providecommand \citenamefont [1]{#1}%
\providecommand\href[0]{\@sanitize\@href}%
\providecommand\@href[1]{\endgroup\@@startlink{#1}\endgroup\@@href}%
\providecommand\@@href[1]{#1\@@endlink}%
\providecommand \@sanitize [0]{\begingroup\catcode`\&12\catcode`\#12\relax}%
\@ifxundefined \pdfoutput {\@firstoftwo}{%
 \@ifnum{\z@=\pdfoutput}{\@firstoftwo}{\@secondoftwo}%
}{%
 \providecommand\@@startlink[1]{\leavevmode\special{html:<a href="#1">}}%
 \providecommand\@@endlink[0]{\special{html:</a>}}%
}{%
 \providecommand\@@startlink[1]{%
  \leavevmode
  \pdfstartlink
   attr{/Border[0 0 1 ]/H/I/C[0 1 1]}%
   user{/Subtype/Link/A<</Type/Action/S/URI/URI(#1)>>}%
  \relax
 }%
 \providecommand\@@endlink[0]{\pdfendlink}%
}%
\providecommand \url  [0]{\begingroup\@sanitize \@url }%
\providecommand \@url [1]{\endgroup\@href {#1}{\urlprefix}}%
\providecommand \urlprefix [0]{URL }%
\providecommand \Eprint[0]{\href }%
\@ifxundefined \urlstyle {%
  \providecommand \doi [1]{doi:\discretionary{}{}{}#1}%
}{%
  \providecommand \doi [0]{doi:\discretionary{}{}{}\begingroup
  \urlstyle{rm}\Url }%
}%
\providecommand \doibase [0]{http://dx.doi.org/}%
\providecommand \Doi[1]{\href{\doibase#1}}%
\providecommand \bibAnnote [3]{%
  \BibitemShut{#1}%
  \begin{quotation}\noindent
    \textsc{Key:}\ #2\\\textsc{Annotation:}\ #3%
  \end{quotation}%
}%
\providecommand \bibAnnoteFile [2]{%
  \IfFileExists{#2}{\bibAnnote {#1} {#2} {\input{#2}}}{}%
}%
\providecommand \typeout [0]{\immediate \write \m@ne }%
\providecommand \selectlanguage [0]{\@gobble}%
\providecommand \bibinfo [0]{\@secondoftwo}%
\providecommand \bibfield [0]{\@secondoftwo}%
\providecommand \translation [1]{[#1]}%
\providecommand \BibitemOpen[0]{}%
\providecommand \bibitemStop [0]{}%
\providecommand \bibitemNoStop [0]{.\EOS\space}%
\providecommand \EOS [0]{\spacefactor3000\relax}%
\providecommand \BibitemShut [1]{\csname bibitem#1\endcsname}%
\bibitem{Susskind:1998vk}%
  \BibitemOpen
  \bibfield{author}{%
  \bibinfo {author} {\bibfnamefont{Leonard}\ \bibnamefont{Susskind}},\ }%
  \bibfield{title}{%
  \enquote{\bibinfo {title} {{Holography in the flat space limit}},}\ }%
  \bibfield{journal}{%
  \Doi{10.1063/1.1301570}{\bibinfo {journal} {AIP Conf. Proc.}}\ }%
  \textbf{\bibinfo {volume} {493}},\ \bibinfo {pages} {98--112} (\bibinfo
  {year} {1999}),\
  \Eprint{http://arxiv.org/abs/hep-th/9901079}{arXiv:hep-th/9901079}%
  \bibAnnoteFile{NoStop}{Susskind:1998vk}%
\bibitem{deBoer:2003vf}%
  \BibitemOpen
  \bibfield{author}{%
  \bibinfo {author} {\bibfnamefont{Jan}\ \bibnamefont{de~Boer}}\ and\ \bibinfo
  {author} {\bibfnamefont{Sergey~N.}\ \bibnamefont{Solodukhin}},\ }%
  \bibfield{title}{%
  \enquote{\bibinfo {title} {{A Holographic reduction of Minkowski
  space-time}},}\ }%
  \bibfield{journal}{%
  \Doi{10.1016/S0550-3213(03)00494-2}{\bibinfo {journal} {Nucl. Phys. B}}\ }%
  \textbf{\bibinfo {volume} {665}},\ \bibinfo {pages} {545--593} (\bibinfo
  {year} {2003}),\
  \Eprint{http://arxiv.org/abs/hep-th/0303006}{arXiv:hep-th/0303006}%
  \bibAnnoteFile{NoStop}{deBoer:2003vf}%
\bibitem{Mann:2005yr}%
  \BibitemOpen
  \bibfield{author}{%
  \bibinfo {author} {\bibfnamefont{Robert~B.}\ \bibnamefont{Mann}}\ and\
  \bibinfo {author} {\bibfnamefont{Donald}\ \bibnamefont{Marolf}},\ }%
  \bibfield{title}{%
  \enquote{\bibinfo {title} {{Holographic renormalization of asymptotically
  flat spacetimes}},}\ }%
  \bibfield{journal}{%
  \Doi{10.1088/0264-9381/23/9/010}{\bibinfo {journal} {Class. Quant. Grav.}}\
  }%
  \textbf{\bibinfo {volume} {23}},\ \bibinfo {pages} {2927--2950} (\bibinfo
  {year} {2006}),\
  \Eprint{http://arxiv.org/abs/hep-th/0511096}{arXiv:hep-th/0511096}%
  \bibAnnoteFile{NoStop}{Mann:2005yr}%
\bibitem{Dappiaggi:2005ci}%
  \BibitemOpen
  \bibfield{author}{%
  \bibinfo {author} {\bibfnamefont{Claudio}\ \bibnamefont{Dappiaggi}}, \bibinfo
  {author} {\bibfnamefont{Valter}\ \bibnamefont{Moretti}},\ and\ \bibinfo
  {author} {\bibfnamefont{Nicola}\ \bibnamefont{Pinamonti}},\ }%
  \bibfield{title}{%
  \enquote{\bibinfo {title} {{Rigorous steps towards holography in
  asymptotically flat spacetimes}},}\ }%
  \bibfield{journal}{%
  \Doi{10.1142/S0129055X0600270X}{\bibinfo {journal} {Rev. Math. Phys.}}\ }%
  \textbf{\bibinfo {volume} {18}},\ \bibinfo {pages} {349--416} (\bibinfo
  {year} {2006}),\
  \Eprint{http://arxiv.org/abs/gr-qc/0506069}{arXiv:gr-qc/0506069}%
  \bibAnnoteFile{NoStop}{Dappiaggi:2005ci}%
\bibitem{Barnich:2010eb}%
  \BibitemOpen
  \bibfield{author}{%
  \bibinfo {author} {\bibfnamefont{Glenn}\ \bibnamefont{Barnich}}\ and\
  \bibinfo {author} {\bibfnamefont{Cedric}\ \bibnamefont{Troessaert}},\ }%
  \bibfield{title}{%
  \enquote{\bibinfo {title} {{Aspects of the BMS/CFT correspondence}},}\ }%
  \bibfield{journal}{%
  \Doi{10.1007/JHEP05(2010)062}{\bibinfo {journal} {JHEP}}\ }%
  \textbf{\bibinfo {volume} {05}},\ \bibinfo {pages} {062} (\bibinfo {year}
  {2010}),\ \Eprint{http://arxiv.org/abs/1001.1541}{arXiv:1001.1541 [hep-th]}%
  \bibAnnoteFile{NoStop}{Barnich:2010eb}%
\bibitem{Bagchi:2010zz}%
  \BibitemOpen
  \bibfield{author}{%
  \bibinfo {author} {\bibfnamefont{Arjun}\ \bibnamefont{Bagchi}},\ }%
  \bibfield{title}{%
  \enquote{\bibinfo {title} {{Correspondence between Asymptotically Flat
  Spacetimes and Nonrelativistic Conformal Field Theories}},}\ }%
  \bibfield{journal}{%
  \Doi{10.1103/PhysRevLett.105.171601}{\bibinfo {journal} {Phys. Rev. Lett.}}\
  }%
  \textbf{\bibinfo {volume} {105}},\ \bibinfo {pages} {171601} (\bibinfo {year}
  {2010}),\ \Eprint{http://arxiv.org/abs/1006.3354}{arXiv:1006.3354 [hep-th]}%
  \bibAnnoteFile{NoStop}{Bagchi:2010zz}%
\bibitem{Banks:2014iha}%
  \BibitemOpen
  \bibfield{author}{%
  \bibinfo {author} {\bibfnamefont{T.}~\bibnamefont{Banks}},\ }%
  \bibfield{title}{%
  \enquote{\bibinfo {title} {{The Super BMS Algebra, Scattering and
  Holography}},}\ }%
   (\bibinfo {month} {3}\ \bibinfo {year} {2014}),\
  \Eprint{http://arxiv.org/abs/1403.3420}{arXiv:1403.3420 [hep-th]}%
  \bibAnnoteFile{NoStop}{Banks:2014iha}%
\bibitem{Pasterski:2017kqt}%
  \BibitemOpen
  \bibfield{author}{%
  \bibinfo {author} {\bibfnamefont{Sabrina}\ \bibnamefont{Pasterski}}\ and\
  \bibinfo {author} {\bibfnamefont{Shu-Heng}\ \bibnamefont{Shao}},\ }%
  \bibfield{title}{%
  \enquote{\bibinfo {title} {{Conformal basis for flat space amplitudes}},}\ }%
  \bibfield{journal}{%
  \Doi{10.1103/PhysRevD.96.065022}{\bibinfo {journal} {Phys. Rev. D}}\ }%
  \textbf{\bibinfo {volume} {96}},\ \bibinfo {pages} {065022} (\bibinfo {year}
  {2017}),\ \Eprint{http://arxiv.org/abs/1705.01027}{arXiv:1705.01027
  [hep-th]}%
  \bibAnnoteFile{NoStop}{Pasterski:2017kqt}%
\bibitem{Kapec:2017gsg}%
  \BibitemOpen
  \bibfield{author}{%
  \bibinfo {author} {\bibfnamefont{Daniel}\ \bibnamefont{Kapec}}\ and\ \bibinfo
  {author} {\bibfnamefont{Prahar}\ \bibnamefont{Mitra}},\ }%
  \bibfield{title}{%
  \enquote{\bibinfo {title} {{A $d$-Dimensional Stress Tensor for Mink$_{d+2}$
  Gravity}},}\ }%
  \bibfield{journal}{%
  \Doi{10.1007/JHEP05(2018)186}{\bibinfo {journal} {JHEP}}\ }%
  \textbf{\bibinfo {volume} {05}},\ \bibinfo {pages} {186} (\bibinfo {year}
  {2018}),\ \Eprint{http://arxiv.org/abs/1711.04371}{arXiv:1711.04371
  [hep-th]}%
  \bibAnnoteFile{NoStop}{Kapec:2017gsg}%
\bibitem{Kapec:2022hih}%
  \BibitemOpen
  \bibfield{author}{%
  \bibinfo {author} {\bibfnamefont{Daniel}\ \bibnamefont{Kapec}},\ }%
  \bibfield{title}{%
  \enquote{\bibinfo {title} {{Soft Particles and Infinite-Dimensional
  Geometry}},}\ }%
   (\bibinfo {month} {10}\ \bibinfo {year} {2022}),\
  \Eprint{http://arxiv.org/abs/2210.00606}{arXiv:2210.00606 [hep-th]}%
  \bibAnnoteFile{NoStop}{Kapec:2022hih}%
\bibitem{Kapec:2022axw}%
  \BibitemOpen
  \bibfield{author}{%
  \bibinfo {author} {\bibfnamefont{Daniel}\ \bibnamefont{Kapec}}, \bibinfo
  {author} {\bibfnamefont{Y.~T.~Albert}\ \bibnamefont{Law}},\ and\ \bibinfo
  {author} {\bibfnamefont{Sruthi~A.}\ \bibnamefont{Narayanan}},\ }%
  \bibfield{title}{%
  \enquote{\bibinfo {title} {{Soft Scalars and the Geometry of the Space of
  Celestial CFTs}},}\ }%
   (\bibinfo {month} {5}\ \bibinfo {year} {2022}),\
  \Eprint{http://arxiv.org/abs/2205.10935}{arXiv:2205.10935 [hep-th]}%
  \bibAnnoteFile{NoStop}{Kapec:2022axw}%
\bibitem{Pasterski:2017ylz}%
  \BibitemOpen
  \bibfield{author}{%
  \bibinfo {author} {\bibfnamefont{Sabrina}\ \bibnamefont{Pasterski}}, \bibinfo
  {author} {\bibfnamefont{Shu-Heng}\ \bibnamefont{Shao}},\ and\ \bibinfo
  {author} {\bibfnamefont{Andrew}\ \bibnamefont{Strominger}},\ }%
  \bibfield{title}{%
  \enquote{\bibinfo {title} {{Gluon Amplitudes as 2d Conformal Correlators}},}\
  }%
  \bibfield{journal}{%
  \Doi{10.1103/PhysRevD.96.085006}{\bibinfo {journal} {Phys. Rev. D}}\ }%
  \textbf{\bibinfo {volume} {96}},\ \bibinfo {pages} {085006} (\bibinfo {year}
  {2017}),\ \Eprint{http://arxiv.org/abs/1706.03917}{arXiv:1706.03917
  [hep-th]}%
  \bibAnnoteFile{NoStop}{Pasterski:2017ylz}%
\bibitem{Schreiber:2017jsr}%
  \BibitemOpen
  \bibfield{author}{%
  \bibinfo {author} {\bibfnamefont{Anders}\ \bibnamefont{Schreiber}}, \bibinfo
  {author} {\bibfnamefont{Anastasia}\ \bibnamefont{Volovich}},\ and\ \bibinfo
  {author} {\bibfnamefont{Michael}\ \bibnamefont{Zlotnikov}},\ }%
  \bibfield{title}{%
  \enquote{\bibinfo {title} {{Tree-level gluon amplitudes on the celestial
  sphere}},}\ }%
  \bibfield{journal}{%
  \Doi{10.1016/j.physletb.2018.04.010}{\bibinfo {journal} {Phys. Lett. B}}\ }%
  \textbf{\bibinfo {volume} {781}},\ \bibinfo {pages} {349--357} (\bibinfo
  {year} {2018}),\ \Eprint{http://arxiv.org/abs/1711.08435}{arXiv:1711.08435
  [hep-th]}%
  \bibAnnoteFile{NoStop}{Schreiber:2017jsr}%
\bibitem{Stieberger:2018edy}%
  \BibitemOpen
  \bibfield{author}{%
  \bibinfo {author} {\bibfnamefont{Stephan}\ \bibnamefont{Stieberger}}\ and\
  \bibinfo {author} {\bibfnamefont{Tomasz~R.}\ \bibnamefont{Taylor}},\ }%
  \bibfield{title}{%
  \enquote{\bibinfo {title} {{Strings on Celestial Sphere}},}\ }%
  \bibfield{journal}{%
  \Doi{10.1016/j.nuclphysb.2018.08.019}{\bibinfo {journal} {Nucl. Phys. B}}\ }%
  \textbf{\bibinfo {volume} {935}},\ \bibinfo {pages} {388--411} (\bibinfo
  {year} {2018}),\ \Eprint{http://arxiv.org/abs/1806.05688}{arXiv:1806.05688
  [hep-th]}%
  \bibAnnoteFile{NoStop}{Stieberger:2018edy}%
\bibitem{Nandan:2019jas}%
  \BibitemOpen
  \bibfield{author}{%
  \bibinfo {author} {\bibfnamefont{Dhritiman}\ \bibnamefont{Nandan}}, \bibinfo
  {author} {\bibfnamefont{Anders}\ \bibnamefont{Schreiber}}, \bibinfo {author}
  {\bibfnamefont{Anastasia}\ \bibnamefont{Volovich}},\ and\ \bibinfo {author}
  {\bibfnamefont{Michael}\ \bibnamefont{Zlotnikov}},\ }%
  \bibfield{title}{%
  \enquote{\bibinfo {title} {{Celestial Amplitudes: Conformal Partial Waves and
  Soft Limits}},}\ }%
  \bibfield{journal}{%
  \Doi{10.1007/JHEP10(2019)018}{\bibinfo {journal} {JHEP}}\ }%
  \textbf{\bibinfo {volume} {10}},\ \bibinfo {pages} {018} (\bibinfo {year}
  {2019}),\ \Eprint{http://arxiv.org/abs/1904.10940}{arXiv:1904.10940
  [hep-th]}%
  \bibAnnoteFile{NoStop}{Nandan:2019jas}%
\bibitem{Pate:2019lpp}%
  \BibitemOpen
  \bibfield{author}{%
  \bibinfo {author} {\bibfnamefont{Monica}\ \bibnamefont{Pate}}, \bibinfo
  {author} {\bibfnamefont{Ana-Maria}\ \bibnamefont{Raclariu}}, \bibinfo
  {author} {\bibfnamefont{Andrew}\ \bibnamefont{Strominger}},\ and\ \bibinfo
  {author} {\bibfnamefont{Ellis~Ye}\ \bibnamefont{Yuan}},\ }%
  \bibfield{title}{%
  \enquote{\bibinfo {title} {{Celestial operator products of gluons and
  gravitons}},}\ }%
  \bibfield{journal}{%
  \Doi{10.1142/S0129055X21400031}{\bibinfo {journal} {Rev. Math. Phys.}}\ }%
  \textbf{\bibinfo {volume} {33}},\ \bibinfo {pages} {2140003} (\bibinfo {year}
  {2021}),\ \Eprint{http://arxiv.org/abs/1910.07424}{arXiv:1910.07424
  [hep-th]}%
  \bibAnnoteFile{NoStop}{Pate:2019lpp}%
\bibitem{Gonzalez:2020tpi}%
  \BibitemOpen
  \bibfield{author}{%
  \bibinfo {author} {\bibfnamefont{Hern\'an~A.}\ \bibnamefont{Gonz\'alez}},
  \bibinfo {author} {\bibfnamefont{Andrea}\ \bibnamefont{Puhm}},\ and\ \bibinfo
  {author} {\bibfnamefont{Francisco}\ \bibnamefont{Rojas}},\ }%
  \bibfield{title}{%
  \enquote{\bibinfo {title} {{Loop corrections to celestial amplitudes}},}\ }%
  \bibfield{journal}{%
  \Doi{10.1103/PhysRevD.102.126027}{\bibinfo {journal} {Phys. Rev. D}}\ }%
  \textbf{\bibinfo {volume} {102}},\ \bibinfo {pages} {126027} (\bibinfo {year}
  {2020}),\ \Eprint{http://arxiv.org/abs/2009.07290}{arXiv:2009.07290
  [hep-th]}%
  \bibAnnoteFile{NoStop}{Gonzalez:2020tpi}%
\bibitem{Banerjee:2020vnt}%
  \BibitemOpen
  \bibfield{author}{%
  \bibinfo {author} {\bibfnamefont{Shamik}\ \bibnamefont{Banerjee}}\ and\
  \bibinfo {author} {\bibfnamefont{Sudip}\ \bibnamefont{Ghosh}},\ }%
  \bibfield{title}{%
  \enquote{\bibinfo {title} {{MHV gluon scattering amplitudes from celestial
  current algebras}},}\ }%
  \bibfield{journal}{%
  \Doi{10.1007/JHEP10(2021)111}{\bibinfo {journal} {JHEP}}\ }%
  \textbf{\bibinfo {volume} {10}},\ \bibinfo {pages} {111} (\bibinfo {year}
  {2021}),\ \Eprint{http://arxiv.org/abs/2011.00017}{arXiv:2011.00017
  [hep-th]}%
  \bibAnnoteFile{NoStop}{Banerjee:2020vnt}%
\bibitem{Arkani-Hamed:2020gyp}%
  \BibitemOpen
  \bibfield{author}{%
  \bibinfo {author} {\bibfnamefont{Nima}\ \bibnamefont{Arkani-Hamed}}, \bibinfo
  {author} {\bibfnamefont{Monica}\ \bibnamefont{Pate}}, \bibinfo {author}
  {\bibfnamefont{Ana-Maria}\ \bibnamefont{Raclariu}},\ and\ \bibinfo {author}
  {\bibfnamefont{Andrew}\ \bibnamefont{Strominger}},\ }%
  \bibfield{title}{%
  \enquote{\bibinfo {title} {{Celestial amplitudes from UV to IR}},}\ }%
  \bibfield{journal}{%
  \Doi{10.1007/JHEP08(2021)062}{\bibinfo {journal} {JHEP}}\ }%
  \textbf{\bibinfo {volume} {08}},\ \bibinfo {pages} {062} (\bibinfo {year}
  {2021}),\ \Eprint{http://arxiv.org/abs/2012.04208}{arXiv:2012.04208
  [hep-th]}%
  \bibAnnoteFile{NoStop}{Arkani-Hamed:2020gyp}%
\bibitem{Guevara:2021abz}%
  \BibitemOpen
  \bibfield{author}{%
  \bibinfo {author} {\bibfnamefont{Alfredo}\ \bibnamefont{Guevara}}, \bibinfo
  {author} {\bibfnamefont{Elizabeth}\ \bibnamefont{Himwich}}, \bibinfo {author}
  {\bibfnamefont{Monica}\ \bibnamefont{Pate}},\ and\ \bibinfo {author}
  {\bibfnamefont{Andrew}\ \bibnamefont{Strominger}},\ }%
  \bibfield{title}{%
  \enquote{\bibinfo {title} {{Holographic symmetry algebras for gauge theory
  and gravity}},}\ }%
  \bibfield{journal}{%
  \Doi{10.1007/JHEP11(2021)152}{\bibinfo {journal} {JHEP}}\ }%
  \textbf{\bibinfo {volume} {11}},\ \bibinfo {pages} {152} (\bibinfo {year}
  {2021}),\ \Eprint{http://arxiv.org/abs/2103.03961}{arXiv:2103.03961
  [hep-th]}%
  \bibAnnoteFile{NoStop}{Guevara:2021abz}%
\bibitem{Fan:2021isc}%
  \BibitemOpen
  \bibfield{author}{%
  \bibinfo {author} {\bibfnamefont{Wei}\ \bibnamefont{Fan}}, \bibinfo {author}
  {\bibfnamefont{Angelos}\ \bibnamefont{Fotopoulos}}, \bibinfo {author}
  {\bibfnamefont{Stephan}\ \bibnamefont{Stieberger}}, \bibinfo {author}
  {\bibfnamefont{Tomasz~R.}\ \bibnamefont{Taylor}},\ and\ \bibinfo {author}
  {\bibfnamefont{Bin}\ \bibnamefont{Zhu}},\ }%
  \bibfield{title}{%
  \enquote{\bibinfo {title} {{Conformal blocks from celestial gluon
  amplitudes}},}\ }%
  \bibfield{journal}{%
  \Doi{10.1007/JHEP05(2021)170}{\bibinfo {journal} {JHEP}}\ }%
  \textbf{\bibinfo {volume} {05}},\ \bibinfo {pages} {170} (\bibinfo {year}
  {2021}),\ \Eprint{http://arxiv.org/abs/2103.04420}{arXiv:2103.04420
  [hep-th]}%
  \bibAnnoteFile{NoStop}{Fan:2021isc}%
\bibitem{Atanasov:2021cje}%
  \BibitemOpen
  \bibfield{author}{%
  \bibinfo {author} {\bibfnamefont{Alexander}\ \bibnamefont{Atanasov}},
  \bibinfo {author} {\bibfnamefont{Walker}\ \bibnamefont{Melton}}, \bibinfo
  {author} {\bibfnamefont{Ana-Maria}\ \bibnamefont{Raclariu}},\ and\ \bibinfo
  {author} {\bibfnamefont{Andrew}\ \bibnamefont{Strominger}},\ }%
  \bibfield{title}{%
  \enquote{\bibinfo {title} {{Conformal block expansion in celestial CFT}},}\
  }%
  \bibfield{journal}{%
  \Doi{10.1103/PhysRevD.104.126033}{\bibinfo {journal} {Phys. Rev. D}}\ }%
  \textbf{\bibinfo {volume} {104}},\ \bibinfo {pages} {126033} (\bibinfo {year}
  {2021}),\ \Eprint{http://arxiv.org/abs/2104.13432}{arXiv:2104.13432
  [hep-th]}%
  \bibAnnoteFile{NoStop}{Atanasov:2021cje}%
\bibitem{Himwich:2021dau}%
  \BibitemOpen
  \bibfield{author}{%
  \bibinfo {author} {\bibfnamefont{Elizabeth}\ \bibnamefont{Himwich}}, \bibinfo
  {author} {\bibfnamefont{Monica}\ \bibnamefont{Pate}},\ and\ \bibinfo {author}
  {\bibfnamefont{Kyle}\ \bibnamefont{Singh}},\ }%
  \bibfield{title}{%
  \enquote{\bibinfo {title} {{Celestial operator product expansions and
  w$_{1+\infty}$ symmetry for all spins}},}\ }%
  \bibfield{journal}{%
  \Doi{10.1007/JHEP01(2022)080}{\bibinfo {journal} {JHEP}}\ }%
  \textbf{\bibinfo {volume} {01}},\ \bibinfo {pages} {080} (\bibinfo {year}
  {2022}),\ \Eprint{http://arxiv.org/abs/2108.07763}{arXiv:2108.07763
  [hep-th]}%
  \bibAnnoteFile{NoStop}{Himwich:2021dau}%
\bibitem{Fan:2021pbp}%
  \BibitemOpen
  \bibfield{author}{%
  \bibinfo {author} {\bibfnamefont{Wei}\ \bibnamefont{Fan}}, \bibinfo {author}
  {\bibfnamefont{Angelos}\ \bibnamefont{Fotopoulos}}, \bibinfo {author}
  {\bibfnamefont{Stephan}\ \bibnamefont{Stieberger}}, \bibinfo {author}
  {\bibfnamefont{Tomasz~R.}\ \bibnamefont{Taylor}},\ and\ \bibinfo {author}
  {\bibfnamefont{Bin}\ \bibnamefont{Zhu}},\ }%
  \bibfield{title}{%
  \enquote{\bibinfo {title} {{Conformal blocks from celestial gluon amplitudes.
  Part II. Single-valued correlators}},}\ }%
  \bibfield{journal}{%
  \Doi{10.1007/JHEP11(2021)179}{\bibinfo {journal} {JHEP}}\ }%
  \textbf{\bibinfo {volume} {11}},\ \bibinfo {pages} {179} (\bibinfo {year}
  {2021}),\ \Eprint{http://arxiv.org/abs/2108.10337}{arXiv:2108.10337
  [hep-th]}%
  \bibAnnoteFile{NoStop}{Fan:2021pbp}%
\bibitem{Adamo:2021zpw}%
  \BibitemOpen
  \bibfield{author}{%
  \bibinfo {author} {\bibfnamefont{Tim}\ \bibnamefont{Adamo}}, \bibinfo
  {author} {\bibfnamefont{Wei}\ \bibnamefont{Bu}}, \bibinfo {author}
  {\bibfnamefont{Eduardo}\ \bibnamefont{Casali}},\ and\ \bibinfo {author}
  {\bibfnamefont{Atul}\ \bibnamefont{Sharma}},\ }%
  \bibfield{title}{%
  \enquote{\bibinfo {title} {{Celestial operator products from the
  worldsheet}},}\ }%
  \bibfield{journal}{%
  \Doi{10.1007/JHEP06(2022)052}{\bibinfo {journal} {JHEP}}\ }%
  \textbf{\bibinfo {volume} {06}},\ \bibinfo {pages} {052} (\bibinfo {year}
  {2022}),\ \Eprint{http://arxiv.org/abs/2111.02279}{arXiv:2111.02279
  [hep-th]}%
  \bibAnnoteFile{NoStop}{Adamo:2021zpw}%
\bibitem{Hu:2022syq}%
  \BibitemOpen
  \bibfield{author}{%
  \bibinfo {author} {\bibfnamefont{Yangrui}\ \bibnamefont{Hu}}, \bibinfo
  {author} {\bibfnamefont{Luke}\ \bibnamefont{Lippstreu}}, \bibinfo {author}
  {\bibfnamefont{Marcus}\ \bibnamefont{Spradlin}}, \bibinfo {author}
  {\bibfnamefont{Akshay~Yelleshpur}\ \bibnamefont{Srikant}},\ and\ \bibinfo
  {author} {\bibfnamefont{Anastasia}\ \bibnamefont{Volovich}},\ }%
  \bibfield{title}{%
  \enquote{\bibinfo {title} {{Four-point correlators of light-ray operators in
  CCFT}},}\ }%
  \bibfield{journal}{%
  \Doi{10.1007/JHEP07(2022)104}{\bibinfo {journal} {JHEP}}\ }%
  \textbf{\bibinfo {volume} {07}},\ \bibinfo {pages} {104} (\bibinfo {year}
  {2022}),\ \Eprint{http://arxiv.org/abs/2203.04255}{arXiv:2203.04255
  [hep-th]}%
  \bibAnnoteFile{NoStop}{Hu:2022syq}%
\bibitem{Garcia-Sepulveda:2022lga}%
  \BibitemOpen
  \bibfield{author}{%
  \bibinfo {author} {\bibfnamefont{Diego}\
  \bibnamefont{Garc\'\i{}a-Sep\'ulveda}}, \bibinfo {author}
  {\bibfnamefont{Alfredo}\ \bibnamefont{Guevara}}, \bibinfo {author}
  {\bibfnamefont{Justin}\ \bibnamefont{Kulp}},\ and\ \bibinfo {author}
  {\bibfnamefont{Jingxiang}\ \bibnamefont{Wu}},\ }%
  \bibfield{title}{%
  \enquote{\bibinfo {title} {{Notes on resonances and unitarity from celestial
  amplitudes}},}\ }%
  \bibfield{journal}{%
  \Doi{10.1007/JHEP09(2022)245}{\bibinfo {journal} {JHEP}}\ }%
  \textbf{\bibinfo {volume} {09}},\ \bibinfo {pages} {245} (\bibinfo {year}
  {2022}),\ \Eprint{http://arxiv.org/abs/2205.14633}{arXiv:2205.14633
  [hep-th]}%
  \bibAnnoteFile{NoStop}{Garcia-Sepulveda:2022lga}%
\bibitem{Ren:2022sws}%
  \BibitemOpen
  \bibfield{author}{%
  \bibinfo {author} {\bibfnamefont{Lecheng}\ \bibnamefont{Ren}}, \bibinfo
  {author} {\bibfnamefont{Marcus}\ \bibnamefont{Spradlin}}, \bibinfo {author}
  {\bibfnamefont{Akshay}\ \bibnamefont{Yelleshpur~Srikant}},\ and\ \bibinfo
  {author} {\bibfnamefont{Anastasia}\ \bibnamefont{Volovich}},\ }%
  \bibfield{title}{%
  \enquote{\bibinfo {title} {{On effective field theories with celestial
  duals}},}\ }%
  \bibfield{journal}{%
  \Doi{10.1007/JHEP08(2022)251}{\bibinfo {journal} {JHEP}}\ }%
  \textbf{\bibinfo {volume} {08}},\ \bibinfo {pages} {251} (\bibinfo {year}
  {2022}),\ \Eprint{http://arxiv.org/abs/2206.08322}{arXiv:2206.08322
  [hep-th]}%
  \bibAnnoteFile{NoStop}{Ren:2022sws}%
\bibitem{Bhardwaj:2022anh}%
  \BibitemOpen
  \bibfield{author}{%
  \bibinfo {author} {\bibfnamefont{Rishabh}\ \bibnamefont{Bhardwaj}}, \bibinfo
  {author} {\bibfnamefont{Luke}\ \bibnamefont{Lippstreu}}, \bibinfo {author}
  {\bibfnamefont{Lecheng}\ \bibnamefont{Ren}}, \bibinfo {author}
  {\bibfnamefont{Marcus}\ \bibnamefont{Spradlin}}, \bibinfo {author}
  {\bibfnamefont{Akshay}\ \bibnamefont{Yelleshpur~Srikant}},\ and\ \bibinfo
  {author} {\bibfnamefont{Anastasia}\ \bibnamefont{Volovich}},\ }%
  \bibfield{title}{%
  \enquote{\bibinfo {title} {{Loop-level gluon OPEs in celestial
  holography}},}\ }%
   (\bibinfo {month} {8}\ \bibinfo {year} {2022}),\
  \Eprint{http://arxiv.org/abs/2208.14416}{arXiv:2208.14416 [hep-th]}%
  \bibAnnoteFile{NoStop}{Bhardwaj:2022anh}%
\bibitem{Stieberger:2022zyk}%
  \BibitemOpen
  \bibfield{author}{%
  \bibinfo {author} {\bibfnamefont{Stephan}\ \bibnamefont{Stieberger}},
  \bibinfo {author} {\bibfnamefont{Tomasz~R.}\ \bibnamefont{Taylor}},\ and\
  \bibinfo {author} {\bibfnamefont{Bin}\ \bibnamefont{Zhu}},\ }%
  \bibfield{title}{%
  \enquote{\bibinfo {title} {{Celestial Liouville Theory for Yang-Mills
  Amplitudes}},}\ }%
   (\bibinfo {month} {9}\ \bibinfo {year} {2022}),\
  \Eprint{http://arxiv.org/abs/2209.02724}{arXiv:2209.02724 [hep-th]}%
  \bibAnnoteFile{NoStop}{Stieberger:2022zyk}%
\bibitem{Costello:2022wso}%
  \BibitemOpen
  \bibfield{author}{%
  \bibinfo {author} {\bibfnamefont{Kevin}\ \bibnamefont{Costello}}\ and\
  \bibinfo {author} {\bibfnamefont{Natalie~M.}\ \bibnamefont{Paquette}},\ }%
  \bibfield{title}{%
  \enquote{\bibinfo {title} {{Celestial holography meets twisted holography: 4d
  amplitudes from chiral correlators}},}\ }%
   (\bibinfo {month} {1}\ \bibinfo {year} {2022}),\
  \Eprint{http://arxiv.org/abs/2201.02595}{arXiv:2201.02595 [hep-th]}%
  \bibAnnoteFile{NoStop}{Costello:2022wso}%
\bibitem{Costello:2022jpg}%
  \BibitemOpen
  \bibfield{author}{%
  \bibinfo {author} {\bibfnamefont{Kevin}\ \bibnamefont{Costello}}, \bibinfo
  {author} {\bibfnamefont{Natalie~M.}\ \bibnamefont{Paquette}},\ and\ \bibinfo
  {author} {\bibfnamefont{Atul}\ \bibnamefont{Sharma}},\ }%
  \bibfield{title}{%
  \enquote{\bibinfo {title} {{Top-down holography in an asymptotically flat
  spacetime}},}\ }%
   (\bibinfo {month} {8}\ \bibinfo {year} {2022}),\
  \Eprint{http://arxiv.org/abs/2208.14233}{arXiv:2208.14233 [hep-th]}%
  \bibAnnoteFile{NoStop}{Costello:2022jpg}%
\bibitem{Ball:2021tmb}%
  \BibitemOpen
  \bibfield{author}{%
  \bibinfo {author} {\bibfnamefont{Adam}\ \bibnamefont{Ball}}, \bibinfo
  {author} {\bibfnamefont{Sruthi~A.}\ \bibnamefont{Narayanan}}, \bibinfo
  {author} {\bibfnamefont{Jakob}\ \bibnamefont{Salzer}},\ and\ \bibinfo
  {author} {\bibfnamefont{Andrew}\ \bibnamefont{Strominger}},\ }%
  \bibfield{title}{%
  \enquote{\bibinfo {title} {{Perturbatively exact w$_{1+\infty}$ asymptotic
  symmetry of quantum self-dual gravity}},}\ }%
  \bibfield{journal}{%
  \Doi{10.1007/JHEP01(2022)114}{\bibinfo {journal} {JHEP}}\ }%
  \textbf{\bibinfo {volume} {01}},\ \bibinfo {pages} {114} (\bibinfo {year}
  {2022}),\ \Eprint{http://arxiv.org/abs/2111.10392}{arXiv:2111.10392
  [hep-th]}%
  \bibAnnoteFile{NoStop}{Ball:2021tmb}%
\bibitem{Duary:2022onm}%
  \BibitemOpen
  \bibfield{author}{%
  \bibinfo {author} {\bibfnamefont{Sarthak}\ \bibnamefont{Duary}},\ }%
  \bibfield{title}{%
  \enquote{\bibinfo {title} {{Celestial amplitude for 2d theory}},}\ }%
   (\bibinfo {month} {9}\ \bibinfo {year} {2022}),\
  \Eprint{http://arxiv.org/abs/2209.02776}{arXiv:2209.02776 [hep-th]}%
  \bibAnnoteFile{NoStop}{Duary:2022onm}%
\bibitem{Coleman:1967ad}%
  \BibitemOpen
  \bibfield{author}{%
  \bibinfo {author} {\bibfnamefont{Sidney~R.}\ \bibnamefont{Coleman}}\ and\
  \bibinfo {author} {\bibfnamefont{J.}~\bibnamefont{Mandula}},\ }%
  \bibfield{title}{%
  \enquote{\bibinfo {title} {{All Possible Symmetries of the S Matrix}},}\ }%
  \bibfield{journal}{%
  \Doi{10.1103/PhysRev.159.1251}{\bibinfo {journal} {Phys. Rev.}}\ }%
  \textbf{\bibinfo {volume} {159}},\ \bibinfo {pages} {1251--1256} (\bibinfo
  {year} {1967})%
  \bibAnnoteFile{NoStop}{Coleman:1967ad}%
\bibitem{Dorey:1996gd}%
  \BibitemOpen
  \bibfield{author}{%
  \bibinfo {author} {\bibfnamefont{P.}~\bibnamefont{Dorey}},\ }%
  \enquote{\bibinfo {title} {{Exact S matrices}},}\ \ (\bibinfo {year} {1996})\
  pp.\ \bibinfo {pages} {85--125},\
  \Eprint{http://arxiv.org/abs/hep-th/9810026}{arXiv:hep-th/9810026}%
  \bibAnnoteFile{NoStop}{Dorey:1996gd}%
\bibitem{Paulos:2016but}%
  \BibitemOpen
  \bibfield{author}{%
  \bibinfo {author} {\bibfnamefont{Miguel~F.}\ \bibnamefont{Paulos}}, \bibinfo
  {author} {\bibfnamefont{Joao}\ \bibnamefont{Penedones}}, \bibinfo {author}
  {\bibfnamefont{Jonathan}\ \bibnamefont{Toledo}}, \bibinfo {author}
  {\bibfnamefont{Balt~C.}\ \bibnamefont{van Rees}},\ and\ \bibinfo {author}
  {\bibfnamefont{Pedro}\ \bibnamefont{Vieira}},\ }%
  \bibfield{title}{%
  \enquote{\bibinfo {title} {{The S-matrix bootstrap II: two dimensional
  amplitudes}},}\ }%
  \bibfield{journal}{%
  \Doi{10.1007/JHEP11(2017)143}{\bibinfo {journal} {JHEP}}\ }%
  \textbf{\bibinfo {volume} {11}},\ \bibinfo {pages} {143} (\bibinfo {year}
  {2017}),\ \Eprint{http://arxiv.org/abs/1607.06110}{arXiv:1607.06110
  [hep-th]}%
  \bibAnnoteFile{NoStop}{Paulos:2016but}%
\bibitem{Braden:1989bu}%
  \BibitemOpen
  \bibfield{author}{%
  \bibinfo {author} {\bibfnamefont{H.~W.}\ \bibnamefont{Braden}}, \bibinfo
  {author} {\bibfnamefont{Edward}\ \bibnamefont{Corrigan}}, \bibinfo {author}
  {\bibfnamefont{P.~E.}\ \bibnamefont{Dorey}},\ and\ \bibinfo {author}
  {\bibfnamefont{R.}~\bibnamefont{Sasaki}},\ }%
  \bibfield{title}{%
  \enquote{\bibinfo {title} {{Affine Toda Field Theory and Exact S
  Matrices}},}\ }%
  \bibfield{journal}{%
  \Doi{10.1016/0550-3213(90)90648-W}{\bibinfo {journal} {Nucl. Phys. B}}\ }%
  \textbf{\bibinfo {volume} {338}},\ \bibinfo {pages} {689--746} (\bibinfo
  {year} {1990})%
  \bibAnnoteFile{NoStop}{Braden:1989bu}%
\bibitem{Fateev:1990hy}%
  \BibitemOpen
  \bibfield{author}{%
  \bibinfo {author} {\bibfnamefont{V.~A.}\ \bibnamefont{Fateev}}\ and\ \bibinfo
  {author} {\bibfnamefont{A.~B.}\ \bibnamefont{Zamolodchikov}},\ }%
  \bibfield{title}{%
  \enquote{\bibinfo {title} {{Conformal field theory and purely elastic S
  matrices}},}\ }%
  \bibfield{journal}{%
  \Doi{10.1142/S0217751X90000477}{\bibinfo {journal} {Int. J. Mod. Phys. A}}\
  }%
  \textbf{\bibinfo {volume} {5}},\ \bibinfo {pages} {1025--1048} (\bibinfo
  {year} {1990})%
  \bibAnnoteFile{NoStop}{Fateev:1990hy}%
\bibitem{Zamolodchikov:1989fp}%
  \BibitemOpen
  \bibfield{author}{%
  \bibinfo {author} {\bibfnamefont{A.~B.}\ \bibnamefont{Zamolodchikov}},\ }%
  \bibfield{title}{%
  \enquote{\bibinfo {title} {{Integrals of Motion and S Matrix of the (Scaled)
  T=T(c) Ising Model with Magnetic Field}},}\ }%
  \bibfield{journal}{%
  \Doi{10.1142/S0217751X8900176X}{\bibinfo {journal} {Int. J. Mod. Phys. A}}\
  }%
  \textbf{\bibinfo {volume} {4}},\ \bibinfo {pages} {4235} (\bibinfo {year}
  {1989})%
  \bibAnnoteFile{NoStop}{Zamolodchikov:1989fp}%
\bibitem{Zamolodchikov:2013ama}%
  \BibitemOpen
  \bibfield{author}{%
  \bibinfo {author} {\bibfnamefont{Alexander}\ \bibnamefont{Zamolodchikov}},\
  }%
  \bibfield{title}{%
  \enquote{\bibinfo {title} {{Ising Spectroscopy II: Particles and poles at $T
  > T_c$}},}\ }%
   (\bibinfo {month} {10}\ \bibinfo {year} {2013}),\
  \Eprint{http://arxiv.org/abs/1310.4821}{arXiv:1310.4821 [hep-th]}%
  \bibAnnoteFile{NoStop}{Zamolodchikov:2013ama}%
\bibitem{Hollowood:1989cg}%
  \BibitemOpen
  \bibfield{author}{%
  \bibinfo {author} {\bibfnamefont{Timothy~J.}\ \bibnamefont{Hollowood}}\ and\
  \bibinfo {author} {\bibfnamefont{Paul}\ \bibnamefont{Mansfield}},\ }%
  \bibfield{title}{%
  \enquote{\bibinfo {title} {{Rational Conformal Field Theories At, and Away
  From, Criticality as Toda Field Theories}},}\ }%
  \bibfield{journal}{%
  \Doi{10.1016/0370-2693(89)90291-8}{\bibinfo {journal} {Phys. Lett. B}}\ }%
  \textbf{\bibinfo {volume} {226}},\ \bibinfo {pages} {73} (\bibinfo {year}
  {1989})%
  \bibAnnoteFile{NoStop}{Hollowood:1989cg}%
\bibitem{Gabai:2019ryw}%
  \BibitemOpen
  \bibfield{author}{%
  \bibinfo {author} {\bibfnamefont{Barak}\ \bibnamefont{Gabai}}\ and\ \bibinfo
  {author} {\bibfnamefont{Xi}~\bibnamefont{Yin}},\ }%
  \bibfield{title}{%
  \enquote{\bibinfo {title} {{On The S-Matrix of Ising Field Theory in Two
  Dimensions}},}\ }%
   (\bibinfo {month} {5}\ \bibinfo {year} {2019}),\
  \Eprint{http://arxiv.org/abs/1905.00710}{arXiv:1905.00710 [hep-th]}%
  \bibAnnoteFile{NoStop}{Gabai:2019ryw}%
\bibitem{Zamolodchikov:1978xm}%
  \BibitemOpen
  \bibfield{author}{%
  \bibinfo {author} {\bibfnamefont{Alexander~B.}\ \bibnamefont{Zamolodchikov}}\
  and\ \bibinfo {author} {\bibfnamefont{Alexei~B.}\
  \bibnamefont{Zamolodchikov}},\ }%
  \bibfield{title}{%
  \enquote{\bibinfo {title} {{Factorized S Matrices in Two-Dimensions as the
  Exact Solutions of Certain Relativistic Quantum Field Models}},}\ }%
  \bibfield{journal}{%
  \Doi{10.1016/0003-4916(79)90391-9}{\bibinfo {journal} {Annals Phys.}}\ }%
  \textbf{\bibinfo {volume} {120}},\ \bibinfo {pages} {253--291} (\bibinfo
  {year} {1979})%
  \bibAnnoteFile{NoStop}{Zamolodchikov:1978xm}%
\bibitem{Klassen:1989ui}%
  \BibitemOpen
  \bibfield{author}{%
  \bibinfo {author} {\bibfnamefont{Timothy~R.}\ \bibnamefont{Klassen}}\ and\
  \bibinfo {author} {\bibfnamefont{Ezer}\ \bibnamefont{Melzer}},\ }%
  \bibfield{title}{%
  \enquote{\bibinfo {title} {{Purely Elastic Scattering Theories and their
  Ultraviolet Limits}},}\ }%
  \bibfield{journal}{%
  \Doi{10.1016/0550-3213(90)90643-R}{\bibinfo {journal} {Nucl. Phys. B}}\ }%
  \textbf{\bibinfo {volume} {338}},\ \bibinfo {pages} {485--528} (\bibinfo
  {year} {1990})%
  \bibAnnoteFile{NoStop}{Klassen:1989ui}%
\bibitem{Dorey:1990xa}%
  \BibitemOpen
  \bibfield{author}{%
  \bibinfo {author} {\bibfnamefont{Patrick}\ \bibnamefont{Dorey}},\ }%
  \bibfield{title}{%
  \enquote{\bibinfo {title} {{Root systems and purely elastic S matrices}},}\
  }%
  \bibfield{journal}{%
  \Doi{10.1016/0550-3213(91)90428-Z}{\bibinfo {journal} {Nucl. Phys. B}}\ }%
  \textbf{\bibinfo {volume} {358}},\ \bibinfo {pages} {654--676} (\bibinfo
  {year} {1991})%
  \bibAnnoteFile{NoStop}{Dorey:1990xa}%
\bibitem{Dorey:1991zp}%
  \BibitemOpen
  \bibfield{author}{%
  \bibinfo {author} {\bibfnamefont{Patrick}\ \bibnamefont{Dorey}},\ }%
  \bibfield{title}{%
  \enquote{\bibinfo {title} {{Root systems and purely elastic S matrices.
  2.}}.}\ }%
  \bibfield{journal}{%
  \Doi{10.1016/0550-3213(92)90407-3}{\bibinfo {journal} {Nucl. Phys. B}}\ }%
  \textbf{\bibinfo {volume} {374}},\ \bibinfo {pages} {741--761} (\bibinfo
  {year} {1992}),\
  \Eprint{http://arxiv.org/abs/hep-th/9110058}{arXiv:hep-th/9110058}%
  \bibAnnoteFile{NoStop}{Dorey:1991zp}%
\bibitem{Zamolodchikov:1989hfa}%
  \BibitemOpen
  \bibfield{author}{%
  \bibinfo {author} {\bibfnamefont{A.~B.}\ \bibnamefont{Zamolodchikov}},\ }%
  \bibfield{title}{%
  \enquote{\bibinfo {title} {{Integrable field theory from conformal field
  theory}},}\ }%
  \bibfield{journal}{%
  \bibinfo {journal} {Adv. Stud. Pure Math.}\ }%
  \textbf{\bibinfo {volume} {19}},\ \bibinfo {pages} {641--674} (\bibinfo
  {year} {1989})%
  \bibAnnoteFile{NoStop}{Zamolodchikov:1989hfa}%
\bibitem{Dubovsky:2013ira}%
  \BibitemOpen
  \bibfield{author}{%
  \bibinfo {author} {\bibfnamefont{Sergei}\ \bibnamefont{Dubovsky}}, \bibinfo
  {author} {\bibfnamefont{Victor}\ \bibnamefont{Gorbenko}},\ and\ \bibinfo
  {author} {\bibfnamefont{Mehrdad}\ \bibnamefont{Mirbabayi}},\ }%
  \bibfield{title}{%
  \enquote{\bibinfo {title} {{Natural Tuning: Towards A Proof of Concept}},}\
  }%
  \bibfield{journal}{%
  \Doi{10.1007/JHEP09(2013)045}{\bibinfo {journal} {JHEP}}\ }%
  \textbf{\bibinfo {volume} {09}},\ \bibinfo {pages} {045} (\bibinfo {year}
  {2013}),\ \Eprint{http://arxiv.org/abs/1305.6939}{arXiv:1305.6939 [hep-th]}%
  \bibAnnoteFile{NoStop}{Dubovsky:2013ira}%
\bibitem{Dubovsky:2017cnj}%
  \BibitemOpen
  \bibfield{author}{%
  \bibinfo {author} {\bibfnamefont{Sergei}\ \bibnamefont{Dubovsky}}, \bibinfo
  {author} {\bibfnamefont{Victor}\ \bibnamefont{Gorbenko}},\ and\ \bibinfo
  {author} {\bibfnamefont{Mehrdad}\ \bibnamefont{Mirbabayi}},\ }%
  \bibfield{title}{%
  \enquote{\bibinfo {title} {{Asymptotic fragility, near AdS$_{2}$ holography
  and $ T\overline{T} $}},}\ }%
  \bibfield{journal}{%
  \Doi{10.1007/JHEP09(2017)136}{\bibinfo {journal} {JHEP}}\ }%
  \textbf{\bibinfo {volume} {09}},\ \bibinfo {pages} {136} (\bibinfo {year}
  {2017}),\ \Eprint{http://arxiv.org/abs/1706.06604}{arXiv:1706.06604
  [hep-th]}%
  \bibAnnoteFile{NoStop}{Dubovsky:2017cnj}%
\bibitem{Fendley:1993jh}%
  \BibitemOpen
  \bibfield{author}{%
  \bibinfo {author} {\bibfnamefont{P.}~\bibnamefont{Fendley}}\ and\ \bibinfo
  {author} {\bibfnamefont{H.}~\bibnamefont{Saleur}},\ }%
  \enquote{\bibinfo {title} {{Massless integrable quantum field theories and
  massless scattering in (1+1)-dimensions}},}\ in\ \emph{\bibinfo {booktitle}
  {Summer School in High-energy Physics and Cosmology}}\ (\bibinfo {year}
  {1993})\ pp.\ \bibinfo {pages} {301--332},\
  \Eprint{http://arxiv.org/abs/hep-th/9310058}{arXiv:hep-th/9310058}%
  \bibAnnoteFile{NoStop}{Fendley:1993jh}%
\bibitem{Dray:1984ha}%
  \BibitemOpen
  \bibfield{author}{%
  \bibinfo {author} {\bibfnamefont{Tevian}\ \bibnamefont{Dray}}\ and\ \bibinfo
  {author} {\bibfnamefont{Gerard}\ \bibnamefont{'t~Hooft}},\ }%
  \bibfield{title}{%
  \enquote{\bibinfo {title} {{The Gravitational Shock Wave of a Massless
  Particle}},}\ }%
  \bibfield{journal}{%
  \Doi{10.1016/0550-3213(85)90525-5}{\bibinfo {journal} {Nucl. Phys. B}}\ }%
  \textbf{\bibinfo {volume} {253}},\ \bibinfo {pages} {173--188} (\bibinfo
  {year} {1985})%
  \bibAnnoteFile{NoStop}{Dray:1984ha}%
\bibitem{PipolodeGioia:2022exe}%
  \BibitemOpen
  \bibfield{author}{%
  \bibinfo {author} {\bibfnamefont{Leonardo}\ \bibnamefont{Pipolode~Gioia}}\
  and\ \bibinfo {author} {\bibfnamefont{Ana-Maria}\ \bibnamefont{Raclariu}},\
  }%
  \bibfield{title}{%
  \enquote{\bibinfo {title} {{Eikonal Approximation in Celestial CFT}},}\ }%
   (\bibinfo {month} {6}\ \bibinfo {year} {2022}),\
  \Eprint{http://arxiv.org/abs/2206.10547}{arXiv:2206.10547 [hep-th]}%
  \bibAnnoteFile{NoStop}{PipolodeGioia:2022exe}%
\bibitem{Shenker:2013pqa}%
  \BibitemOpen
  \bibfield{author}{%
  \bibinfo {author} {\bibfnamefont{Stephen~H.}\ \bibnamefont{Shenker}}\ and\
  \bibinfo {author} {\bibfnamefont{Douglas}\ \bibnamefont{Stanford}},\ }%
  \bibfield{title}{%
  \enquote{\bibinfo {title} {{Black holes and the butterfly effect}},}\ }%
  \bibfield{journal}{%
  \Doi{10.1007/JHEP03(2014)067}{\bibinfo {journal} {JHEP}}\ }%
  \textbf{\bibinfo {volume} {03}},\ \bibinfo {pages} {067} (\bibinfo {year}
  {2014}),\ \Eprint{http://arxiv.org/abs/1306.0622}{arXiv:1306.0622 [hep-th]}%
  \bibAnnoteFile{NoStop}{Shenker:2013pqa}%
\bibitem{Polchinski:2015cea}%
  \BibitemOpen
  \bibfield{author}{%
  \bibinfo {author} {\bibfnamefont{Joseph}\ \bibnamefont{Polchinski}},\ }%
  \bibfield{title}{%
  \enquote{\bibinfo {title} {{Chaos in the black hole S-matrix}},}\ }%
   (\bibinfo {month} {5}\ \bibinfo {year} {2015}),\
  \Eprint{http://arxiv.org/abs/1505.08108}{arXiv:1505.08108 [hep-th]}%
  \bibAnnoteFile{NoStop}{Polchinski:2015cea}%
\end{thebibliography}%
\end{document}